\documentclass[useAMS,usenatbib,usegraphicx]{mn2e}
\usepackage{subfigure}      
\title[]{The Diversity of Planetary Systems Architectures: Contrasting Theory with Observations}

\author[Y. Miguel, O. M. Guilera and A. Brunini]{Y. Miguel$^{1,2}$\thanks{E-mail: ymiguel@fcaglp.unlp.edu.ar}, O. M. Guilera$^{1,2}$ and A. Brunini$^{1,2}$\thanks{Member of the Carrera del Investigador Cient\'\i fico. Consejo Nacional de Investigaciones Cient\'\i ficas y T\'ecnicas (CONICET).E-mail: abrunini@fcaglp.unlp.edu.ar}\\
$^1$Facultad de Ciencias Astron\'omicas y Geof\'\i sicas. Universidad
Nacional de La Plata. Paseo del Bosque s/n, La Plata (1900), Argentina.\\
$^2$Instituto de Astrof\'\i sica de La Plata (CCT La Plata-CONICET, UNLP), Paseo del Bosque s/n, La Plata (1900), Argentina}

\begin{document}  

\pagerange{\pageref{firstpage}--\pageref{lastpage}}

\label{firstpage}

\maketitle
        
\begin{abstract}

In order to explain the observed diversity of planetary systems architectures
and relate this primordial diversity with the initial properties of the disc
where they were born, we develop a semi-analytical model for computing
planetary system formation. The model is based on the core instability model
for the gas accretion of the embryos and the oligarchic growth regime for the
accretion of the solid cores. Two regimes of planetary migration are also
included. With this model, we consider different initial conditions based on
recent results in protoplanetary discs observations, to generate a variety of
planetary systems. These systems are analyzed statistically, exploring the
importance of several factors that define the planetary systems birth
environment. We explore the relevance of the mass and size of the disc,
metallicity, mass of the central star and time-scale of gaseous disc
dissipation, in defining the architecture of the planetary system. We also
test different values of some key parameters of our model, to find out which
factors best reproduce the diverse sample of observed planetary systems. We
assume different migration rates and initial disc profiles, in the context of
a surface density profile motivated by similarity solutions. According to
this, and based on recent protoplanetary discs observational data, we predict
which systems are the most common in the solar neighbourhood. We intend to
unveil, whether our Solar System is a rarity or more planetary systems like
our own are expected to be found in the near future. We also analyze which is
the more favourable environment for the formation of habitable planets. Our
results show that planetary systems with only terrestrial planets are the most
common, being the only planetary systems formed when considering low
metallicity discs and which also represent the best environment for the
developing of rocky, potentially habitable planets. We also found that
planetary systems like our own are not rare in the solar neighbourhood, being
its formation favoured in massive discs where there is not a large
accumulation of solids in the inner region of the disc. Regarding the
planetary systems that harbor hot and warm Jupiter planets, we found that this systems
are born in very massive, metal-rich discs. Also a fast migration rate is
required in order to form these systems. According to our results, most of the
hot and warm Jupiter systems are composed by only one giant planet, which is also a tendency of the current observational data.

\end{abstract}


\section{Introduction}

Up to date, the set of planetary systems discovered orbiting around single stars, similar to the sun, in the solar neighborhood, ascends to 315 (http://exoplanets.org/), of which 237 are apparently single-planet systems, while the remaining 78 are multiple-planet systems. The first multiple planetary system discovered orbiting a single star, was the one around $47~Uma$, which up to date harbor two confirmed planets of masses $2.5$ and $0.5$ Jupiter masses ($M_{J}$) and semi-major axis $2.1$ and $3.6~au$, respectively \citep{b70,b60} and one inferred planet of $1.6~M_{J}$ located at $11.6~au$ from the central star \citep{b62}, which means that this planetary system host three giant planets located at distances greater than $1~au$ from the central star. On the other extreme, another example is the system $GJ~876$, which houses four planets (two giant planets, one Neptune and one super-earth) located at distances less than $1~au$ \citep{b63,b67,b64,b65,b66}. There are also examples of systems with a hot-Jupiter and another giant planet, located away from the central star at a distance greater than $1~au$, as the system $HD~217107$ \citep{b68,b69}. As seen in these examples, the planetary systems population is remarkably diverse. It displays a wide range of architectures with properties that reflects the environment where they were born and the different mechanisms of formation and evolution and are of special interest to test theoretical models of planetary system formation. 

 All the information provided by the observations of planetary systems has not still been analyzed by theoretical models, although in recent years there have been a few works dealing with planetary systems formation and evolution, that intend to explain some of the observed trends of planetary systems. Such is the work of \citet{b56}, who present self consistent numerical simulations of planetary systems formation and study specifically how the properties of a mature planetary system map to those of its birth disc. \citet{b61} developed a semi-analytical code for planetary systems formation, where they include the effect of resonant capture between embryos during type I migration and the calculation of embryos� orbital and mass evolution after the gas depletion. Their aim was to show that the formation of super-Earths close to the star is possible and that they are more common than hot-Jupiters. Nevertheless, questions as: how common are the planetary systems like our own in the solar neighborhood?, what factors influence the architecture of planetary systems?, what are the differences and similarities between planetary systems? and which is diversity of planetary systems expected to be find in the solar neighborhood?, remain uncertain.

With these questions in mind, is our main objective to explore the importance of several factors in defining the architecture of a planetary system. We also intend to explain the observed diversity of planetary systems and link them to their birth environment. We explore different gas and solids disc profiles, as well as different planetary migration rates, to find out which factors reproduce the different planetary systems observed. According to this, and based on the protoplanetary discs observational data, we predict the systems that will be the more common and thus those expected to be found in the solar neighborhood. In this way, we intend to unveil, whether our Solar System is a rarity or more planetary systems like our own are expected to be found in the near future. We also analyze which is the more favourable environment for the formation of habitable planets.

Based on the most accepted scenario for explaining the formation of planetary systems, our semi-analytical model adopts the core instability model for giant planet formation. In this scenario, the mass distribution in the protoplanetary disc is important, since it defines the number and location of the final giant planets that would define the architecture of the planetary system. So we consider a large range of discs sizes and masses according to recent protoplanetary discs observations \citep{b2,b3} and assume a nebula with a surface density profile motivated by similarity solutions for viscous accretion discs \citep{b4,b5}, which intend to be simple but is a computationally feasible description, consistent with protoplanetary discs observations. Our embryos start growing in the oligarchic growth regime \citep{b12} and if they are able to accrete gas, their gaseous envelope will grow according to a prescription found from results obtained by \citet{b19}. As the embryos are embedded in a gaseous disc, we also include the effects of their mutual interaction, considering type I and II regimes of planetary migration.

We found that those planetary systems that host small rocky planets are the
most common in the solar neighbourhood. These ``low mass planet systems'' are
the only ones that form in a low metallicity environment and represent the
best site for the formation and developing of terrestrial planets in the
habitable zone. The final number of embryos that harbor these planetary
systems is strongly dependent on the initial disc profile and migration rate
assumed. Another striking result is that planetary systems similar to our
Solar System are expected to be common in the solar neighbourhood. These
systems are formed in massive discs, whith no preferential areas for the
accumulation of solids. Finally, we found that those planetary systems with
only hot ($a<0.07~au$) and warm ($0.07<a<1~au$) Jupiter planets, need a very massive, metal-rich disc to be formed and also a fast migration rate. Our results are consistent with the observational trend and show that most of these systems harbor only one giant planet.

\section{Description of the Model} \label{modelo}

We have developed a semi analytical model for planetary systems formation. This model was explained in detail in our previous work \citep{b11}, in this section we will summarize it for completeness. 

\subsection{Protoplanetary Disc}\label{nebulosa}

The minimum mass solar nebula model (MMSN) of \citet{b1} is usually used for
modeling the protoplanetary nebula. As was explained in previous works
\citep{b36,b37,b11} this model suffers from multiple limitations. In order to
avoid these limitations, we adopt a different model to represent the initial
protoplanetary disc structure. Following \citet{b2} , we assume that
the gaseous surfece density of the disc is,

\begin{equation}\label{dengas}
\Sigma_g(a)=\Sigma_g^0 \bigg(\frac{a}{a_c}\bigg)^{-\gamma} e^{-\big(\frac{a}{a_c}\big)^{2-\gamma}}
\end{equation}   
expression based on the work of \citet{b4,b5}. In the equation, $a_c$ is a
parameter introduced to smoothly end the disc and is fitted from the
observations, the $\gamma$ exponent indicates how the material is distributed
on the disc and $\Sigma_g^0$ is a constant value, which is calculated from the
disc's mass and also depends on the characteristic radius, and on the adopted
disc profile.

In a similar way the solid surface density distribution, $\Sigma_s(a)$, is 

\begin{equation}
\Sigma_s(a)=\Sigma_s^0 \eta_{ice}\bigg(\frac{a}{a_c}\bigg)^{-\gamma} e^{-\big(\frac{a}{a_c}\big)^{2-\gamma}}
\end{equation}   
with $\eta_{ice}$ a function which is $1/4$ inside the snow line and $1$ outside it, representing the change in the solids due to water condensation (\citet{b1,b90,b57}).
 
We note that the relation between the gas and solids surface density distribution gives the abundance of heavy elements. For the case of a disc orbiting a star with metallicity $[Fe/H]$, it is \citep{b120,b54, b57},

\begin{equation}\label{relacion-metalicidades}
\bigg(\frac{\Sigma_s^0}{\Sigma_g^0}\bigg)_{\star}= \bigg(\frac{\Sigma_s^0}{\Sigma_g^0}\bigg)_{\odot}10^{[Fe/H]}= z_0 10^{[Fe/H]}
\end{equation}
where $z_0$ is the primordial abundance of heavy elements in the Sun that was found to be $z_0=0.0149$ by \citet{b6}. In this work we will assume that the stellar metallicities follow a log-normal distribution fitted from the results of the CORALIE sample \citep{b51}.

Recent protoplanetary discs observations show that the exponent in the inner part of the disc ($\gamma$) takes values between $0.4$ and $1.1$ \citep{b2}. Following these observations and with the end of comparing with the minimum mass solar nebula case, we explore three different values for this exponent: $\gamma= 0.5,1$ and $1.5$. 

The sample of discs generated, have masses that follows a Log-Gaussian distribution fitted from recent protoplanetary discs observations \citep{b2,b3}.

Since gravitational instabilities can occur in any region of the disc if it becomes cool enough or the mass distribution is really high, then we check the stability of the discs generated. In the linear regime, the gravitational instability limit is given by the Toomre-parameter \citep{b39},

\begin{equation}\label{q0}
Q=\frac{c_s ~k}{\pi~G~\Sigma_g}
\end{equation}
with $k$ the epicyclic frequency of the disc. When the discs are Keplerian, then $k=\Omega_k$ and the Q parameter is

\begin{equation}\label{q}
Q \simeq 1.24 \times 10^5 \bigg(\frac{a}{1au}\bigg)^{\gamma-\frac{7}{4}}\bigg(\frac{a_c}{1au}\bigg)^{-\gamma}\bigg(\frac{M_{\star}}{M_{\odot}}\bigg)\frac{e^{(\frac{a}{a_c})^{2-\gamma}}}{\Sigma_g^0}
\end{equation}
where a value of $Q \leq 1$ represents an unstable disc. 

As shown in the equation, the disc stability depends on the initial disc profile, so discs with the same mass could be stable or unstable depending on the value of $\gamma$ adopted. Those discs characterized by $\gamma=0.5$ are much more massive in the outer disc than those characterized by larger values of $\gamma$. As a consequence, discs with relatively small mass and $\gamma=0.5$ could develop gravitational instabilities, while discs with higher values of $\gamma$ require a larger disc mass in order to undergo gravitational instabilities. As a result, we found that when the highest value of $\gamma$ is considered ($\gamma=1.5$), there are discs with masses up to $1~M_{\odot}$ that present vaues of $Q>1$ in the entire disc. These extremely massive discs should not be consider as keplerian. In order to avoid these discs, we add another condition for the stability. We assume that a disc is stable if $Q>1$ and its mass do not exceed the $20\%$ of the mass of the central star \citep{b5,b85}.

The location of the inner boundary of the dust disc was found by \citet{b21}, through observations of young stellar objects and is,

\begin{equation}\label{innerradius}
a_{in}=0.0688 \bigg(\frac{1500^{\circ}K}{T_{sub}}\bigg)^2\bigg(\frac{L_{\star}}{L_{\odot}}\bigg)^{\frac{1}{2}} au
\end{equation}
with $T_{sub}$ the dust sublimation temperature taken as $1500^{\circ}K$ and $L_{\star}$ and $L_{\odot}$ are the stellar and Sun luminosity respectively. Though \citet{b21} found that this is the inner radius of the dust disc, we adopt this value to represent the end of both discs.   

We locate one initial embryo at the inner radius and the others are separated a distance $\Delta a$ from each other until the end of the disc is reached. This outer edge is the one that contains $95\%$ of the total disc mass, so it is not always the same and as result the initial number of embryos, $N_{ini}$, will be different according to the disc initial properties. As was shown in our previous work \citep{b11}, the greater the mass of the disc, the lower the initial number of embryos. This is because the separation between the embryos of mass $M_t$ is 

\begin{equation}
\Delta a= 10 \bigg(\frac{2M_t}{3M_{\star}}\bigg)^{\frac{1}{3}}a
\end{equation}
this separation is greater as the larger is the initial mass of the embryo and as a result masive discs have a less $N_{ini}$.

Finally our discs are not time invariant. We model the evolution of the gaseous disc with a very simple exponential decay model, which empties the gaseous disc everywhere in time-scales between $10^6$ and $10^7$ years in agreement to observation of circumstellar discs \citep{b41,b42}. The discs of solids change locally due to the accretion of the embryos.

\subsection{The Growth of the Embryos}

In the beginning there are $N_{ini}$ initial embryos embedded in a swarm of planetesimals in a gaseous disc. These embryos will grow due to the accretion of solids, gas and also due to the merger with other embryos. 

The relative velocity between the embryo and the neighbors planetesimals is an important factor in determining different regimes of embryo growth. Our model begins when the cores have enough mass for increasing the velocity dispersion of the surrounding planetesimals, fact that leads to a slow regime that will dominate the embryo growth. This is the oligarchic growth regime \citep{b12}.

The initial mass necessary for exciting the neighbor planetesimals and turn on the oligarchic growth regime was found by \citet{b13} and is given by,

\begin{equation}\label{Masa-inicial}
M_{oli}\simeq \frac{1.6 a^{\frac{6}{5}}10^{\frac{3}{5}}m^{\frac{3}{5}}\Sigma_s^{\frac{3}{5}}}{M_{\star}^{\frac{1}{5}}}
\end{equation}
with m the effective planetesimal mass.  

The solid accretion rate in this regime of growth depends on the radius ($R_p$) and total mass of the planet and the velocity dispersion ($\sigma$) which in turn depends on the planetesimals' eccentricity. Following \citet{b14} the embryo eats planetesimals on a rate: 

\begin{equation}
\frac{dM_{s}}{dt}=10.53 \Sigma_s \, \Omega \, R_p^2\bigg(1+\frac{2GM_t}{R_p\sigma}\bigg)
\end{equation}
where $\Omega$ is the Kepler frequency. As the solid accretion rate depends on
the planetesimals' eccentricity, we assume that they reached an equilibrium
value due to the balance between the protoplanets' gravitational perturbations
and the gas drag effect \citep{b15}.  The accretion of solids ends when the
solid surface density is equal to zero, it means that the embryos consume most
of the planetesimals available in their feeding zone and scattered the others \citep{b15,b16}. 

Up to now we assume that the only mechanism for the growth of the embryos is
the accretion of small planetesimals, but they also can grow due to collisions
with other embryos. This is a very important effect that suffer most of the
embryos \citep{b10} and determines their final characteristics. The giant
impacts between embryos result in the merger of the cores and therefore
represent a sudden and big increase in the mass of the embryo. We assume that
when two cores are located at a distance less than 3.5 Hill radius they will
collide and merge \citep{b17,b110}. 

Following \citet{b17}, we consider that when the embryos have enough mass to be able to retain a gaseous atmosphere, this gas increases the collision cross section of the embryo and the solid accretion rate is enhanced. At first, this gaseous envelope is able to maintain hydrostatic equilibrium, but when the core reaches a critical mass, the envelope can no longer be maintain by hydrostatic equilibrium and the gas accretion onto the core begins \citep{b18,b40}. Following \citet{b16} we assume a simplified formula for this critical mass given by,

\begin{equation} \label{mcritica}
M_{crit}\sim 10 \bigg(\frac{\dot{M_c}}{10^{-6}M_{\oplus}yr^{-1}}\bigg)^{\frac{1}{4}}
\end{equation} 

The gaseous envelope contracts on its own, on a time-scale given by the
following formula that we obtained by fitting the results of the self-consistent code develop by \citet{b19},

\begin{equation} \label{acregas1}
\frac{dM_g}{dt}=\frac{M_t}{\tau_g}
\end{equation}
where $M_g$ is the mass of the surrounding envelope and $\tau_g$ is its characteristic Kelvin-Helmholtz growth time-scale given by, 

\begin{equation} \label{acregas2}
\tau_g=8.35 \times 10^{10} \bigg(\frac{M_t}{M_{\oplus}}\bigg)^{-4.89}yrs
\end{equation}

\subsection{Orbital Evolution of the Embryos}\label{migracion}

While the embryos are managed to form, their are embedded in a gaseous disc which interacts with the cores, leading to a orbital evolution of the embryos called planetary migration.

There are different regimes of planetary migration. We assume that our cores migrate due to type I and II planetary migration as is explained in the following.

\paragraph{Migration Type I.}
This regime acts on low mass planets, which are treated as a small perturbation and the linearized hydrodynamic equations are solved for the disc response. This regime was studied by \citet{b22,b26} and leads to an orbital motion of the embryos towards the central star. Following \citet{b23} the migration rate is,

\begin{equation}\label{migI}
\bigg(\frac{da}{dt}\bigg)_{migI}= c_{migI}[2.7+1.1\beta] \bigg(\frac{M_t}{M_{\star}}\bigg) \frac{\Sigma_g\, a^2}{M_{\star}}\bigg(\frac{a\Omega_K}{c_s}\bigg)^2 a\, \Omega_K 
\end{equation} 
where 
\begin{equation}
\beta=-\frac{d\, log(\Sigma_g)}{d\, log(a)}=\gamma+(2-\gamma)\bigg(\frac{a}{a_c}\bigg)^{2-\gamma}
\end{equation} 
as the time scale for type I migration can be shorter than the disc lifetime, the factor $c_{migI}$ is introduced for considering effects that might stop or slow down migration (e.g.,\citet{b27,b34,b35,b36,b83,b84}) without introducing a mayor degree of complexity to the model.

\paragraph{Migration Type II}
When we are in the presence of a very massive embryo the problem can no longer
be treated as linear and then the disc response should be treated as a non
linear case. This is the type II migration regime \citep{b24,b25}, which for a
disc with a viscosity characterized by $\alpha=10^{-3}$, is \citep{b16},                           

\begin{displaymath}
\bigg(\frac{da}{dt}\bigg)_{migII}\simeq 3sign(a-R_m)\alpha \frac{\Sigma_g(R_m)R_m^2}{M_t}\, \frac{\Omega_K(R_m)}{\Omega_K}
\end{displaymath}
\begin{equation}
\bigg(\frac{h(R_m)}{a}\bigg)^2\, a\Omega_K(R_m) 
\end{equation}  
where,
\begin{equation}
R_m=10 e^{\frac{2t}{\tau_{disc}}}~au
\end{equation}
and $\tau_{disc}$ is the gaseous disc depletion time-scale.

The embryos stop migrating when they reach the inner edge of the disc.

\section{Results}\label{resultados}

Our main objective is to know which is the typical composition and architecture that is expected to be found in a planetary system. To accomplish this goal and following earlier work of \citet{b16,b9,b57,b11}, we performed a series of Monte Carlo numerical simulations. We assume different unknown parameters, such as the density profile and the type I migration rate, in order to compare with the observations and find out which suits better the observational sample of planetary systems.  
 
Following recent results in observations in protoplanetary discs \citep{b2,b3}, we assumed that the exponent which characterizes the distribution of mass in the inner part of the disc adopts three values: $\gamma=0.5,\, 1$ and $1.5$. On the other hand we suppose that the type I migration rate can be slow down 10 and 100 times, but we also analyze the cases when it is not delayed and when planetary migration is not considered.   

Furthermore, we are also interested in finding out which are the parameters that link a planetary system with its birth disc and define the planetary system main characteristics. In order to study this problem, in each simulation we generate 1000 planetary systems, where the initial conditions for each birth disc are taken random from:

\begin{itemize}
\item{}The time-scale for the gas depletion, has a uniform log distribution between $10^6$ and $10^7$ years.
\item{}The stellar mass has a uniform distribution in log scale in the range of $0.7-1.4M_{\odot}$.
\item{}The distribution of metallicities of solar-like stars in the solar neighborhood follows a Gaussian distribution with $\mu=-0.02$ and dispersion 0.22 \citep{b57}.
\item{}The total mass of the disc is well approximated by a log-Gaussian distribution with mean $-2.05$ and dispersion $0.85$. We obtained this value by assuming a log-Gaussian distribution and performed a non-linear least square fit to the sample observed by \citet{b2}  and \citet{b3}.
\item{}The characteristic radius, $a_c$ is also well approximated by a log-Gaussian distribution with $\mu=3.8$ and $\sigma=0.18$. This distribution was obtained with the same procedure described in the previous item.
\end{itemize}

We want to know how many of these planetary systems generated in our simulations match with an observed one. So, with the aim of comparing with the observations
, we assumed that an observed planetary system matches quantitatively with an artificial one, when the masses and semi-major axis of its giant planets are the same to less than 10\%. Up to date, there are 315 planetary systems found orbiting a single star, similar to the Sun, 66\% of them could be reproduced quantitatively by our simulations. The remaining 33\% are multiple planetary systems, where we could not find artificial systems whose planets match exactly the mass and semi-major axis than observed, although we have found qualitatively similar systems, that will be shown in the next section, where we also analyze the different architectures found, the characteristics of the simulated planetary systems, and how this final characteristics map the discs where they were born.   

\subsection{A New Classification for Planetary Systems}

So far, more than 300 planetary systems have been found orbiting a single star. These planetary systems present different characteristics and with the end of understanding their formation, composition and relation with their birth disc, it is appropriate to classify them according to their architecture. 

Since most of the observed planets are giant planets, we use them for the
planetary systems classification. In order to determine which mass is the
appropriate for separating planets into giant planets and low mass planets, we look from the theoretical point of view. Planets with masses larger than $\sim 15~M_{\oplus}$ have reached the crossover mass, which means that the runaway gas accretion process has began. This process ends only when there is no residual gas in the disc or a gap form near planet's orbit, so planets with masses larger than $15~M_{\oplus}$ are giant planets or failed giant planets (=Neptunes). These are the planets considered for our classification.

On the other hand, based on the observations, we note that these planets are located either near the central star (at distances less than $1~au$), in an intermediate zone or far away (at a distances larger than $30~au$), a fact that is the basis of our classification.

Then we separate all the planetary systems according to the following classification:

\begin{itemize}
\item \textbf{``hot and warm Jupiter systems''}: these planetary systems host planets with masses larger than $15~M_{\oplus}$ at a distance less than $1~au$.
\item \textbf{``solar systems''}: a planetary system is an analog of our Solar System if it harbors giant planets or Neptunes located between $1$ and $30~au$. 
\item \textbf{``combined systems''}: these planetary systems harbor at least one giant planet within $1~au$ and at least one in the middle part of the disc, between $1$ and $30~au$.  
\item \textbf{``cold-Jupiter systems''}: if the giant planets are located further from $30~au$ then it is a cold-Jupiter system.   
\item \textbf{``low mass planet systems''}: these systems have only planets with masses less than $15~M_{\oplus}$.
\end{itemize}

In the table \ref{estadistica-observados} we show the statistics of the population of planetary systems observed. In order to compare with our population of artificial planetary systems we eliminate of the observational sample those planetary systems formed around binary or multiple stars, as well as those where the mass of the central star is less than $0.7$ or greater than $1.4 M_{\odot}$. The total number of observed planetary systems analyzed in the table is 315\footnote{http://exoplanets.org/}. 

\begin{table*}
 \centering
 \begin{minipage}{105mm}
\caption{Percentage (\%) of the different planetary systems detected, whose planets orbit around a single star with mass in the range of $0.7$-$1.4 M_{\odot}$.}
\begin{tabular}{lc}
\hline
Type of Planetary System & Percentage of Planetary Systems Detected (\%)\\
\hline
\hline
Hot and warm Jupiters & 52.38 \\
Solar systems & 31.43 \\
Cold Jupiters & 0 \\
Combined systems & 16.19 \\
Low mass planet systems  & 0 \\
\hline
\label{estadistica-observados}
\end{tabular}
\end{minipage}
\end{table*}

We note that most of the planetary systems are hot and warm Jupiters systems, but we also found a large percentage of planetary systems analogs to the Solar System, although these do not harbor planets like the Earth, or these were not detected yet. It must be noticed that this sample is biased towards those planets that are easier to be detected with the current observational techniques, but we hope that in the near future planets like our own will be easier to be detected and this will improve the statistics, but in the meantime some predictions can be made about what we hope to find. 

To this end, in the following we will explore the statistics of the population
of planetary systems found in our simulations. Tables \ref{estadisticagama05}, \ref{estadisticagama1}  and \ref{estadisticagama15}, show the percentage of different kinds of planetary systems found when assuming different initial disc profiles: $\gamma=0.5$,$1$ and $1.5$ respectively. In each table, we show the statistics when different migration rates are considered. We perform simulations without migration and adopting different type I migration rates, which are indicated in the different columns. 

\textit{The first striking result is that in all cases, speaking of all discs ($\gamma=0.5$, $1$ or $1.5$) and all the migration rates (without migration and when $C_{migI}=0.01$, $0.1$ and $1$), it is always the planetary systems with small planets which are the vast majority.} This means that these systems are the invisible majority, because none of them has been detected yet. But if the observational techniques allowed it, a lot of them would be detected and this is what we expect to find in the close future. 

We also noticed that we found zero cold-Jupiters systems in all the analyzed cases. This is because the core instability model allows the formation of giant planets close to the star, being the snow line the preferred zone \citep{b16}, which is believed to be initially located in the inner part of the disc. This model for planetary formation can not explain the formation of planets as far as $30~au$. Another possibility could be that the planet migrates outward, which could occur if it shares a resonance with another giant planet and the inner one is significantly more massive than the outer one \citep{b94}. Finally, we should consider the possibility that the planet may have formed in the inner planetary system and then ejected outwards. Nevertheless, none of these hypotheses are addressed in our study. Therefore, we need a different scenario to explain the origin of these giant planets, as could be a different mechanism of formation (see e.g. \citet{b20,b28}), or migration, or consider the formation and subsequent ejection of the planets towards the outer system  \citep{b91,b92,b93}.

Finally we point out that as we note in tables \ref{estadisticagama05}, \ref{estadisticagama1}  and \ref{estadisticagama15}, there are failed planetary systems. These systems were born in very low mass discs, which did not allow the formation of objects with masses larger than planet Mercury's mass.

\begin{table*}
 \centering
 \begin{minipage}{120mm}
\caption{Percentage (\%) of planetary systems formed when $\gamma=0.5$ and different migration rates.}
\begin{tabular}{lcccc}
\hline
\multicolumn{5}{|c|}{$\gamma=0.5$}\\
\hline
Type of Planetary System & No migration & $C_{migI}=0.01$  & $C_{migI}=0.1$  & $C_{migI}=1$ \\
\hline
\hline
Hot and warm Jupiters & 0.2 & 2.7 & 4.3 & 4.1 \\
Solar systems & 11.3 & 12.6 & 9.1 & 1.8 \\
Cold Jupiter & 0 & 0 & 0 & 0 0\\
Combined systems & 0 & 3 & 2.2 & 0.5 \\
Low mass planet systems & 77.2 & 70.8 & 73.4 & 81.9\\
Failed planetary systems & 11.3 & 10.9 & 11 & 11.7 \\
\hline
\label{estadisticagama05}
\end{tabular}
\end{minipage}
\end{table*}

Table \ref{estadisticagama05} shows the numerical results when $\gamma=0.5$
and, as seen in the table, when planetary migration is not considered we found
a large percentage of systems analogs to our Solar System, while the
percentage of hot and warm Jupiter Systems is really small. This is because giant planets are formed in regions of higher accumulation of solids. The solids surface density profile considered in this case, allows the formation of these planets only near the ice line, since there is no accumulation of material in other regions of the disc. In addition, if the migration is not considered, they stay were they were formed.

When the migration is considered the planets have a radial motion that moved
them towards the star and, as a consequence, the number of hot and warm Jupiters and
combined systems increases. We also note that when migration is the fastest
($c_{migI}=1$), the population of low mass planet systems increases, this is because in this case the time-scale for type I migration is really fast and inhibits the growth of embryos.

Table \ref{estadisticagama1} shows the statistics when we assume a surface
density profile with an exponent that characterizes the inner part of the disc
equal to $1$. Assuming a sharper disc profile, implies that the solids are
accumulated in the inner regions of the disc. This accumulation allows the
formation of a larger number of giant planets and as a result the number of
rocky and failed systems decrease, as seen in the table, in all the cases with the exception of the case where the parameter for delaying type I migration is equal to 1. As was said, this faster case inhibits the growth of the embryos, even for a $\gamma=1$ profile.

\begin{table*}
 \centering
 \begin{minipage}{120mm}
\caption{Percentage (\%) of planetary systems formed when $\gamma=1$ and different values of $C_{migI}$.}
\begin{tabular}{lcccc}
\hline
\multicolumn{5}{|c|}{$\gamma=1$}\\
\hline
Type of Planetary System & No migration & $C_{migI}=0.01$  & $C_{migI}=0.1$  & $C_{migI}=1$ \\
\hline
\hline
Hot and warm Jupiters & 1.8 & 8.2 & 11.1 & 5.8 \\
Solar systems & 23.7 & 19.9 & 7.7 & 1.4 \\
Cold Jupiters & 0 & 0 & 0 & 0 0\\
Combined systems & 0 & 9.3 & 6.3 & 0.3 \\
Low mass planet systems & 73.4 & 61.6 & 72.8 & 88.3\\
Failed planetary systems & 1.1 & 1 & 2.1 & 4.2 \\
\hline
\label{estadisticagama1}
\end{tabular}
\end{minipage}
\end{table*}

When we assume a profile with $\gamma=1.5$, we found the results shown in
table \ref{estadisticagama15}. This case is characterized by a large
accumulation of solids in the inner disc and a lower density in the outer
region. Then the population of low mass planet systems is still decreasing and the
number of hot and warm Jupiter systems is still increasing, because the solids
surface density is very high in the inner part of the disc, while it falls
rapidly beyond the snow line, fact that favors the formation of Jupiter
planets inside 1~au, and as a consequence we found a larger percentage of hot and warm Jupiter systems in this case.

\begin{table*}
 \centering
 \begin{minipage}{120mm}
\caption{Percentage (\%) of planetary systems formed when $\gamma=1.5$ and different values of $C_{migI}$.}
\begin{tabular}{lcccc}
\hline
\multicolumn{5}{|c|}{$\gamma=1.5$}\\
\hline
Type of Planetary System & No migration & $C_{migI}=0.01$  & $C_{migI}=0.1$  & $C_{migI}=1$ \\
\hline
\hline
Hot and warm Jupiter & 4.5 & 15.3 & 14.8 & 3.7 \\
Solar systems & 27.1 & 16.4 & 7.2 & 0 \\
Cold Jupiters & 0 & 0 & 0 & 0\\
Combined systems& 0.9 & 15.6 & 6 & 0 \\
Low mass planet systems & 67.3 & 52.5 & 64.6 & 56.6\\
Failed planetary systems & 0.2 & 0.2 & 7.4 & 39.7 \\
\hline
\label{estadisticagama15}
\end{tabular}
\end{minipage}
\end{table*}

\subsection{Characterizing Different Types of Planetary Systems}

We divided the planetary systems in five classes and showed a statistical overview of them, comparing with the observational data. In this section we explore in detail each class of planetary system. 

\subsubsection{Hot and Warm Jupiter Systems}

As seen in table \ref{estadistica-observados}, 52.38 \% of the planetary
systems known so far are hot and warm Jupiter systems. This is because the hot-Jupiters are easier to be detected, but according to our results this type of planetary systems are not the most common in the solar neighbourhood. 

\begin{figure*}
  \begin{minipage}{120mm}
    \begin{center}
      \includegraphics[angle=270,width=1.\textwidth]{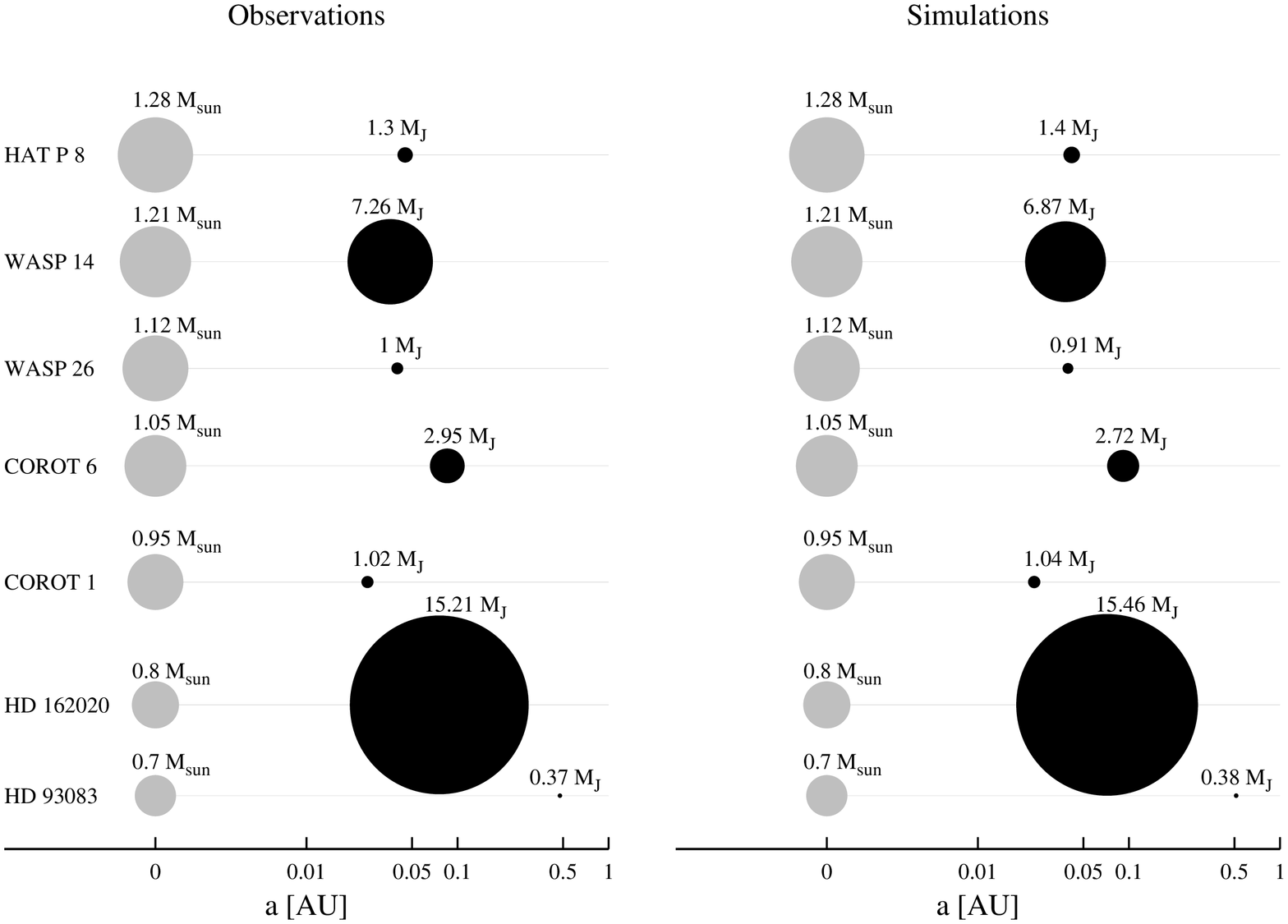}
    \end{center}
    \caption{The figure shows some examples of hot and warm Jupiter systems
      with a single-planet observed, shown in the first column, compared to
      similar hot and warm Jupiter systems generated in our simulations (second column). The semi-major axis is shown in the abscissas, while the size of the circle indicates the planet's mass, which is also printed in Jupiter masses ($M_{J}$) above each planet.}
    \label{compara-hj}
  \end{minipage}
\end{figure*}

Figure \ref{compara-hj} shows a comparison between some examples of
hot and warm Jupiter systems detected and some systems artificially formed with our
model. In the figure, we compare only the characteristics of the planets in the systems, not the stars. Also, when speaking of artificial systems, only the planets
with masses larger than $15~M_{\oplus}$ are plotted.  This is because terrestrial planets evolve for $\simeq 100$ to $200~Myr$. During this time the gravitational interactions between the embryos (after nebular gas was dissipated) play a fundamental role. In this work we only considered the evolution of a system during 20 Myr. and we did not take into account the gravitational interactions between the embryos. For these reasons, we consider that our results can only be compared with observations of giant exoplanets.

As seen in the figure,
our model reproduce quantitatively the hot and warm Jupiter systems observed.

The observed systems HAT-P-8 (first row), WASP-26 (third row) and HD 162020 (sixth row), are similar to artificial systems that were generated assuming a disc profile characterized by $\gamma=1$ and the fastest migration rate. The artificial systems generated in the second and fifth row, were formed considering a disc profile with $\gamma=1.5$ and also $c_{migI}=1$. The system analog to Corot 6 (fourth row) was formed when assuming $\gamma=1.5$ and a migration rate delayed only 10 times and finally, the system with the smallest hot-Jupiter shown in last row, was formed under the assumption of a disc profile characterized by $\gamma=1.5$ and no migration. 

As seen in the figures, with the exception of the last system, the others were formed assuming a disc profile characterized by a large accumulation of solids in the inner disc and the fastest migration rate assumed in this work. These are the preferred conditions for the formation of these systems. The artificial system generated in the seventh row, is a rare system, that was formed in a very massive disc where the solids are abundant in the inner disc ($\gamma=1.5$) and as a consequence a hot-Jupiter in situ was allowed to form. 

An analysis of observational data shows that there are 315 extrasolar
planetary systems observed, of which 165 harbor giant planets located inside
1~au. Of these 165, 17 ($\sim 10$~\%) correspond to multiple systems while the remain are single planetary systems ($\sim 90$~\%). In order to determine
whether this is a feature of these systems, we analyze the number of giant
planets that is expected to be found in a system of this kind. 

\begin{figure*}
  \begin{minipage}{120mm}
    \begin{center}
      \includegraphics[angle=270,width=1.\textwidth]{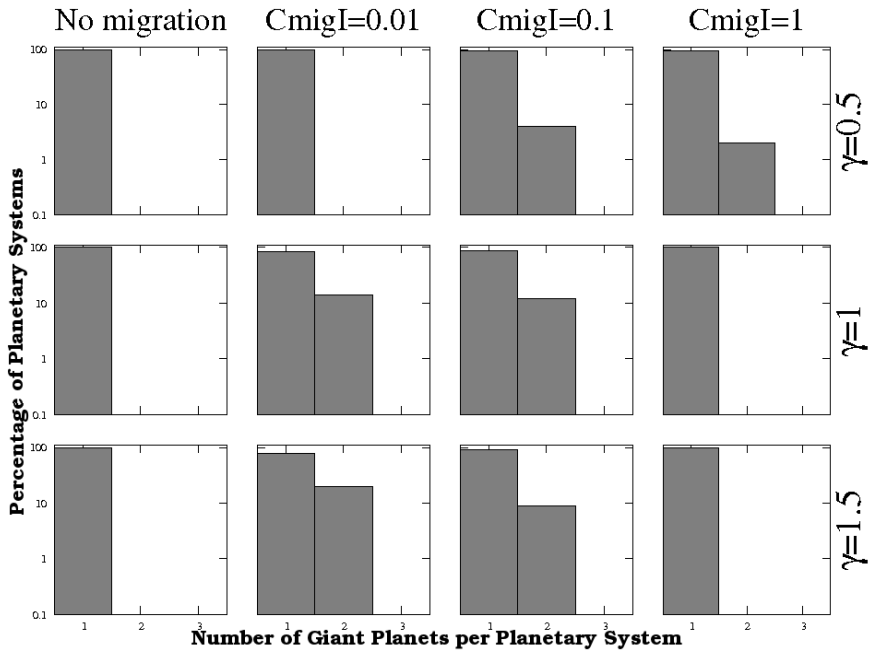}
    \end{center}
    \caption{Histograms showing the number of giant planets per hot and warm jupiter
      system. In the figure, the number of planets with masses larger than
      $15~M_{\oplus}$ per planetary system is shown in the x-axis and the
      y-axis shows the percentage of hot and warm jupiter systems, which is shown in log-scale. The different rows show the resulting histograms for different values of $\gamma$ and in the columns the results of assuming different type I migration rates are shown. }
    \label{histo-HJ}
  \end{minipage}
\end{figure*}

Figure \ref{histo-HJ} shows histograms reproducing the percentage of
artificial hot and warm Jupiter systems that harbor one, two or three giant
planets. Hot and warm Jupiter systems with more than 3 planets were not formed. The different rows show the results when different values of $\gamma$ are considered. In the first row $\gamma=0.5$ , in the second $\gamma=1$ and the last one shows the resulting histograms when $\gamma=1.5$. Different columns show the results when different parameters for delaying type I migration are considered. In the first column the migration was not considered, in the second column $c_{migI}=0.01$, in the third column $c_{migI}=0.1$ and the last case shows the histograms when migration is not delayed. 

We note that the number of giant planets that harbor a hot and warm Jupiter
system is strongly dependent on the type I migration rate assumed. When the
migration is not considered (first column), all the hot and warm Jupiter systems harbor only one giant planet. These are planets that were formed in situ and, as seen in the previous section, these systems are very rare. 

When migration is considered but delayed 100 times (second column), there are
hot and warm Jupiter systems which harbor two giant planets, but only those systems
that were formed assuming an steeper disc profile ($\gamma=1$ or $1.5$). A
disc characterized by a profile with an exponent in the inner part given by
$\gamma=1$ or $\gamma=1.5$ , is a disc that favors the formation of giant
planets and because planets form closer to the star than when $\gamma=0.5$,
the migration push them towards the star and the system becomes a hot and warm
Jupiter system.  

In the third column the migration is delayed only 10 times, and we note that
when the migration is faster, all the explored disc profiles formed hot and
warm Jupiter systems with two giant planets.     

Finally, the last column shows the results when the migration is not
reduced. We note that $\gamma=0.5$ is the only case when two giant
planets are allowed to form. This is because the giant planets are formed
further from the central star than in the other cases, and although the fast migration inhibits the growth,
they could grow by colliding with other embryos on its path. Nevertheless, in
this case the percentage of planetary systems with two giant planets is really small, less
than $5\%$. Then a fast migration rate does not favour the formation of several
giant planets either. 

We see that a migration so fast inhibits the embryo's growth, which is why
several giant planets in the same system are not allowed to form and as a
result all the hot and warm Jupiter systems host only one giant planet.

As a general conclusion, we note that in all cases analyzed, most of the hot
and warm Jupiter systems are composed by only one giant planet, which is also a tendency of the current observational data \citep{b110}. 

\subsubsection{Solar Systems}

So far, 99 of the observed planetary systems around single stars, could be classified as solar systems. These 99 systems represent $\sim 31.5 \%$ of all the planetary systems found, as was shown in table \ref{estadistica-observados}. 

\begin{figure*}
  \begin{minipage}{120mm}
    \begin{center}
      \includegraphics[angle=270,width=1.\textwidth]{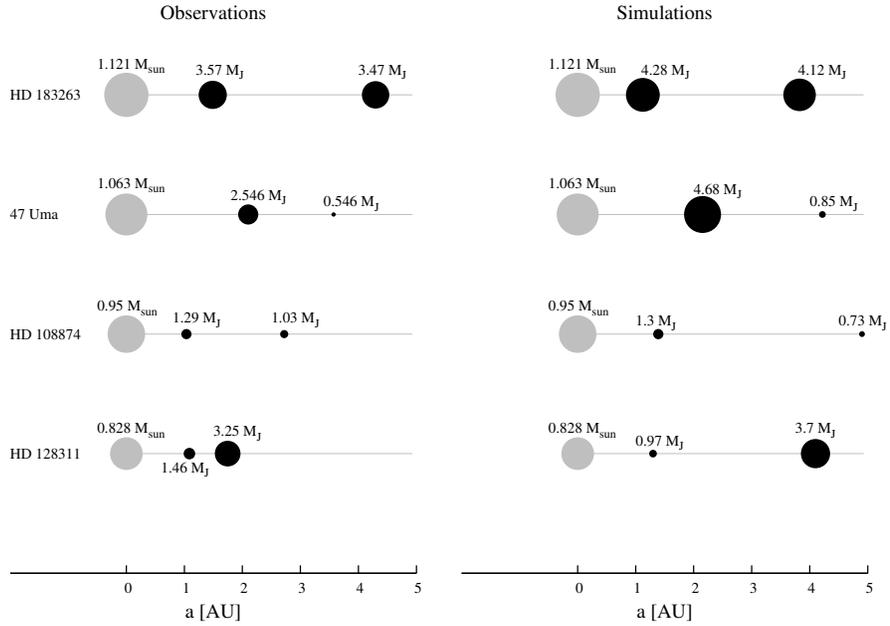}
    \end{center}
    \caption{A sample of multiple solar systems observed, shown in the first column, compared to some examples generated in our simulations (second column). Positions of a circle along the x-axis indicate the planets location, while the size of the circle indicates their mass. The mass of the planet in $M_J$ , is indicated above each planet.}
    \label{compara-ss}
  \end{minipage}
\end{figure*}

Figure \ref{compara-ss} shows a qualitative comparison between some examples of observed solar systems, which are shown in the first column and where only the giant planets are plotted, and some generated with our simulations, shown in the second column.  In the figure, the semi-major axis of the planets are indicated in the x-axis, while their mass is represented with the different sizes of the black circles and printed above each planet. As seen in the figure the observational sample of solar systems can be reproduced qualitatively by our simulations.  

In the figure, the artificial solar system shown in the first row, was formed considering that $\gamma=1$ and $c_{migI}=0.1$. In the second row, the solar system was generated assuming  $\gamma=0.5$ and $c_{migI}=0.1$. As seen in both cases, a flat disc is needed, in order to form these systems. This is because a small value of $\gamma$, favors the formation of several giant planets further from the central star. If a solar system was born from a disc with a great accumulation in the inner disc, as is the case when $\gamma=1.5$, which is also the case most similar to the minimum mass solar nebula model of \citet{b1}, a very slow or zero migration is required in order to the giant planets formed in the ice line to stay there and do not become hot-Jupiters. That is precisely what is observed in the artificial system formed in the third row, which was generated assuming a disc profile characterized by $\gamma=1.5$ and where the planetary migration was not considered. Finally, the planetary system formed in the fourth row, was born in a disc characterized by $\gamma=1$ and where the embryos do not migrate, which is consistent with the preferred scenario for the formation of these systems. 

\begin{figure*}
  \begin{minipage}{120mm}
    \begin{center}
      \includegraphics[angle=270,width=1.\textwidth]{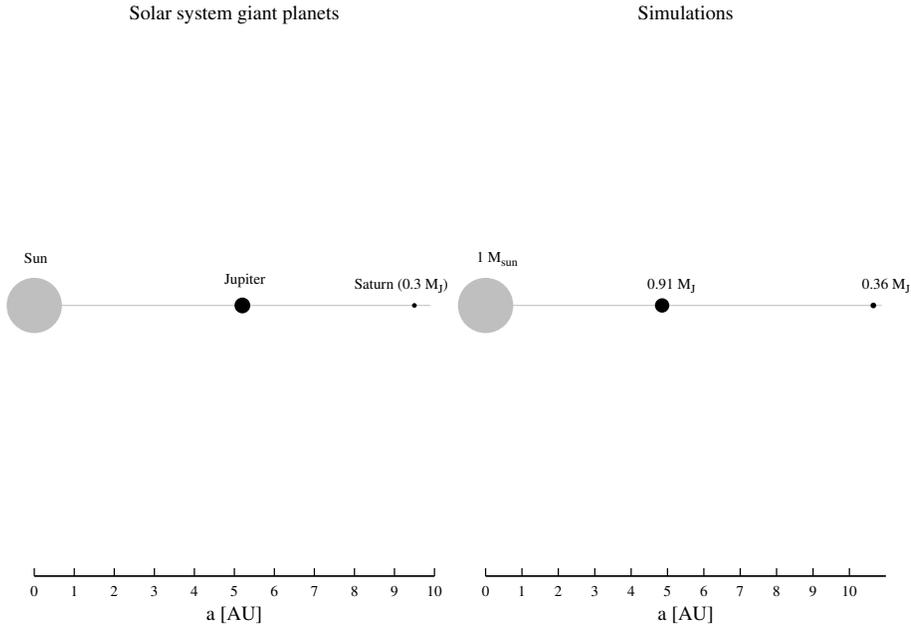}
    \end{center}
    \caption{The figure shows a comparison between our Solar System (first column) and an artificial one generated with our simulations (second column). The semi-major axis of the planets is shown along the x-axis, while the size of the circle indicates their mass. The mass of the planets is also indicated in Jupiter's mass.}
    \label{sistemasolar}
  \end{minipage}
\end{figure*}

We can not ignore our own Solar System, so a qualitatively comparison of our
planetary system compared to a generated one is shown in Figure
\ref{sistemasolar}. The distance between the planet and its central star is
shown in the x-axis and its mass is shown with the different size of the
circle and it is also written for each planet. 

In the Figure only Jupiter and Saturn are shown, because we do not found any
artificial planetary system with the exact mass and location of all the giant
planets in our Solar System. Then, the solar system shown in the figure,
harbors only two giant planets that match with our Solar System. For the
formation of the artificial solar system we assume that  $\gamma=0.5$ and the
planetary migration is not considered. As was said, a system with a very large
value of $\gamma$ do not favor the formation of several giant planets in the
same system and located further away from the central star. On the other hand
a faster migration rate would locate the planets in the inner disc and the
system would became a hot and warm Jupiter system. So a small value of $\gamma$ and a slow migration rate are the best conditions when trying to reproduce our Solar System.

Looking at the architecture of the 100 solar systems detected so far (where we include our Solar System), it can be noticed that only five of them have more than one giant planet, in addition, and with the exception of our Solar System, all of them harbor two giant planets. Since we do not know if this sample is representative of all solar systems or is due to an observational bias, we analyze statistically our numerical results.

\begin{figure*}
  \begin{minipage}{120mm}
    \begin{center}
      \includegraphics[angle=270,width=1.\textwidth]{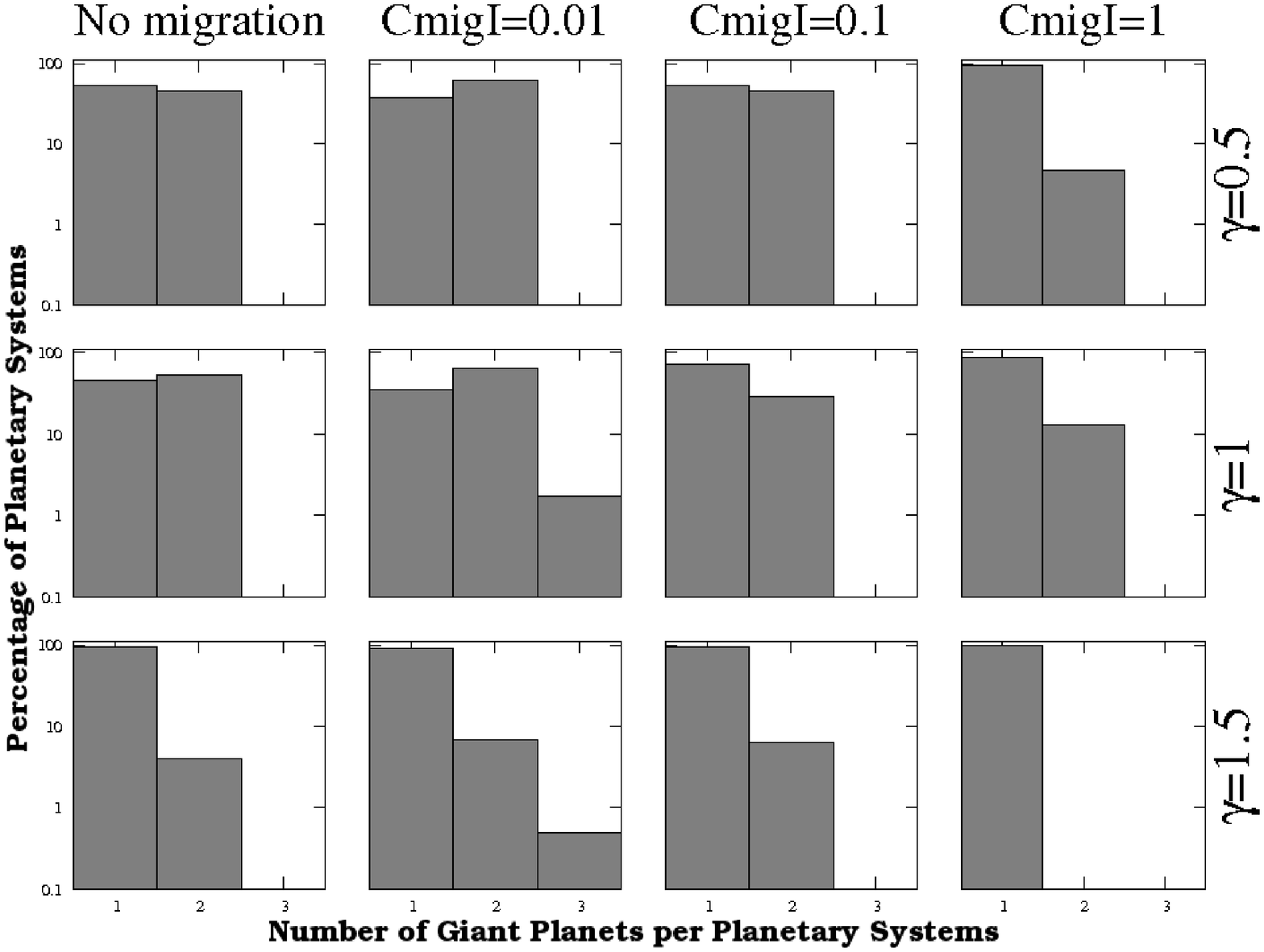}
    \end{center}
    \caption{The figure shows histograms with the percentage of solar systems with 1, 2 or 3 planets with masses larger than $15~M_{\oplus}$. We present the results found with all the analyzed cases, where the rows show the results found when different initial disc profiles are assumed ($\gamma=0.5$, $1$ and $1.5$) and the columns show the numerical results when we assume different migration rates, $c_{migI}=0$, $0.01$, $0.1$ and $1$.}
    \label{histo-SS}
  \end{minipage}
\end{figure*}

Figure \ref{histo-SS} displays histograms showing the percentage of solar
systems that harbor one, two or three planets with masses larger than
$15M_{\oplus}$. This figure is analog to the figure \ref{histo-HJ}, where the columns
show the results when  diferent migration rates are considered and the three
rows show the results when different disc profiles are assumed.  

The figure shows that solar systems with 2 giant planets are pretty
common and those with three giant planets are very rare. This is because
according to the core instability model, to form two giant planets, a large
amount of solid material is needed. This allow the rapid formation of very
massive cores, which start the gas accretion, becoming a giant planets. This
usually occurs in the snow line region, which is the preference formation zone for these planets. If the disc is very massive and has a high
metallicity, another giant planet could be formed, which does not ocurr in
most cases. 

The first row shows the results when the initial density profile of the disc
is characterized by an exponent of $\gamma=0.5$.  The density profile in these
discs is very smooth and there is no accumulation of gas and solids in the
inner disc. As a consequence, the formation of giant planets may occur farther
from the central star than in the case of steeper profiles. The migration move
them towards the star but it is not enough to locate them inside $1~au$ and
for these reason these systems remain as solar systems. Since in this case the
formation of several planets further from the central star is favored, in some
cases there are more solar systems with two giant planets, than those formed
with a sinlge one. 

When the density profile is a bit steeper, $\gamma=1$, we note that, there are
still several planetary systems with two giant planets, even with three in
some few cases, so this disc profile also allows the formation of several giant planets per disc.

Finally, in the case of the steepest profile ($\gamma=1.5$) we see that the
overall percentage of systems with multiple giant planets have fallen compared
with previous cases, because a larger amount of solids in the
inner part of the disc promotess the formation of a single giant planet per
disc \citep{b7}.

\subsubsection{Combined Systems}

These systems represent an intermediate class between the hot and warm Jupiter systems
and those analogous to the Solar System. Since belonging to this class implies
the presence of at least one planet within $1\, au$ and at least one outside,
all these systems have two or more planets with masses greater than
$15~M_{\oplus}$. Our results agree with the observations regarding the
frequency of these systems, although it could be a coincidence due to an observational bias. According to the observations, $\sim 16$~\% of planetary systems discovered so far belong to this class, while our simulations show that in no case combined systems generated exceed the $\sim 15$~\% of the artificial sample.

\begin{figure*}
  \begin{minipage}{120mm}
    \begin{center}
      \includegraphics[angle=270,width=1.\textwidth]{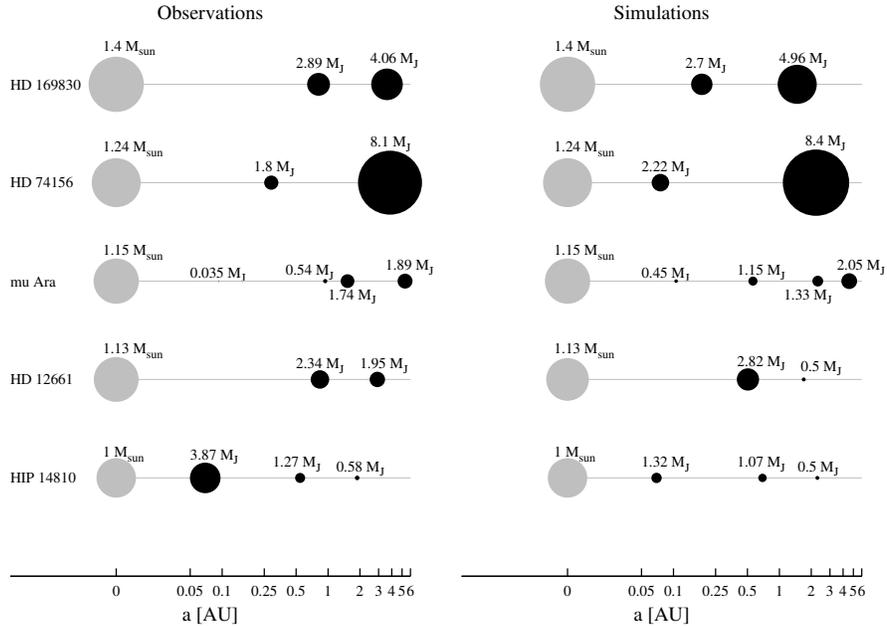}
    \end{center}
    \caption{Examples of planetary systems observed (first column), compared to those generated in our simulations (second column), where only the giant planets were plotted. Positions of a circle along the abscissa indicate the planets semi-major axis. The size of the circle indicates the planetary mass. In addition, the mass of the planet (in $M_{J}$) , is printed above each planet.}
    \label{mixtos-cualitativo}
  \end{minipage}
\end{figure*}

Figure \ref{mixtos-cualitativo} is analogous to figure \ref{compara-hj}, but
in this case we show a comparison of combined systems. 

As seen in the Figure, the combined systems generated in our simulations match
qualitatively with the observed systems, so that the architecture of these
systems can be explained by the core accretion model and the current models of
planetary migration. It is also noticed that while these systems belong to
the same class, they are qualitatively different, because multiple factors
determine the final architecture of a planetary system. Some of these factors
are the initial mass of the disc, the metallicty, the gaseous disipation
time-scale, the distribution of gas and solids along the disc and the
migration time scale. 

The first row shows a system with two giant planets in which the mass of the
inner planet is smaller than the mass of the exterior one. The artificial
system was created with a disc characterized by a profile with $\gamma=1$ and
a slow migration rate ($c_{migI} = 0.01$). This system was formed in a disc
with a smooth profile, where the more massive planet was formed in the region
of greatest accumulation of solids (the snow line), while the lower mass planet
 was formed closer to the central star, inside the snow line. Since the migration was delayed 100 times, these planets did not move too far from the region where they were born. 

We also found systems such as the one shown in the second row, where the mass
of the planets also grows outward but in this case the difference between the
mass of the inner and outer planet is much greater than in the previous
case. The artificial system was created with a profile characterized by
$\gamma=1.5$ and slow migration ($c_{migI}=0.01$). This disc profile is
characterized by a great accumulation of solids in inner disc, specially at
the snow line, causing the rapid formation of the most massive planet in this
region. The lower mass planet was formed closer to the central star and as the profile allows a higher concentration of solids in the interior compared to the case of lower $\gamma$ profiles, migration is also faster, so this planet migrate towards the star, on its path it collide with other embryos and increase its core's mass, being able to accrete gas before it is depleted and form a large gaseous envelope.

The planetary systems shown in the third row also shows that the mass grows
outward, but in this case the systems harbor 4 planets with masses larger than
$15~M_{\oplus}$ and all of them have masses less than $\simeq 2~M_{J}$. The simulated system was generated assuming that $\gamma=1$ and a migration rate faster than in previous cases ($c_{migI}=0.1$). The flat disc profile allows the formation of several giant planets in the same disc, but the growth is slower with this disc profile and the planets have smaller masses when compared to the previous case.

Finally we found systems as those shown in rows 4 and 5, where in both cases
the mass grows inwards and the artificial systems where formed assuming a disc
profile of $\gamma=1$ and a faster migration rate of $c_{migI}=0.1$.  In the
case of planetary system generated in row 4, the most massive planet was born
at the snow line and migrates inwards and the less massive planet was formed
in outer regions with less gas and solids available to accrete. The system
generated in the last row is a system where the embryos acquire most of its cores' mass due to collisions with other embryos. Then the more massive planet migrates until the end of the disc and accrete more solids than the others. 

We also noticed that in rows 3 and 5, the systems were formed under
identical parameters ($\gamma=1$, $c_{migI}=0.1$ ), but they have an opposite
correlation of mass and distance. This implies that other factors must lead to
this difference. The sinthetic system in row 3 was formed in a disc with
$M_d=0.08~M_{\odot}$, $a_c=65~au$ and metellicity of $0.2$. The system formed
in row 5 was born in a $0.11~M_{\odot}$ disc, $a_c=47.85~au$ and
metallicity$=0.026$. This means, that the system in row 5 had more solids and
gas available to form the giant planets. In this masive disc, the migration
rate is faster (because of the larger amount of solids) and therefore the
inner giant planet formed in the ice line and then migrated quickly, eating
the other planets that might have formed there. In the meantime, the other
giant planets formed farther from the star and migrated towards regions with a
major solid surface density, growing in the process. On the other hand, the
system in row 3 have a lower disc mass and most important, the characteristic
radius is much larger as in the previous case. This implies that the solids
are spread over a much larger area and as a consecuence, the planetary
migration was not so fast and the planets remained where they were born. As a
conclusion we note that while the initial disc profile and the migration rate
have a strong influence on the final architecture of the system, there are
also other important parameters in defining it. This will be studied in section \ref{definiendo}.  

\subsubsection{Low Mass Planet Systems}

Low mass planet systems are those who does not host planets with masses larger than
 $15~M_{\oplus}$. Although none of these systems was observed yet, as seen in tables \ref{estadisticagama05}, \ref{estadisticagama1} and \ref{estadisticagama15}, the majority of stars would not host giant planets, so this kind of planetary systems represent a substantial fraction of planetary systems which were not deeply studied yet.

In the standard scenario of rocky planets formation, the finally stage is the
giant impact regime. With our model we are able to study the firsts stages of rocky planets' growth in the context of a disc where several cores are formed simultaneously.  Studying the last stage of rocky planets growth, implies a dynamic monitoring of phenomena such as resonant interactions important during last stages of their growth (eg., \citet{b58,b59}), which are not analyzed with our model. Nevertheless, we believe that our studies are an important complement to numerical dynamical studies that analyze last phases of growth, since we provide the initial conditions for these studies.

With the aim of analyzing the final number of planets found in a low mass planet system as a function of the initial mass of the protoplanetary disc ($M_d$), we divide the planetary systems in three types and analyze the final number of planets found per system in each class. 

The first group are those low mass planet systems formed in low mass discs ($M_d
<0.05~M_{\odot} $), which are the most common low mass planet systems. The second group
are those formed in intermediate mass discs ($0.05~M_{\odot} \le M_d
<0.1~M_{\odot}$) and the last group are the ones originated in very massive
discs ($0.1~M_{\odot}\le M_d$), which are the less common systems, that are
characterised by small metallicities. These low mass planet systems formed in very massive
discs are most common when the migration rate is faster, because this inhibit
the growth of the embryos.

\begin{figure}
  \begin{center}
    \subfigure[]{\label{histo-pch-gama05}\includegraphics[angle=270,width=.5\textwidth]{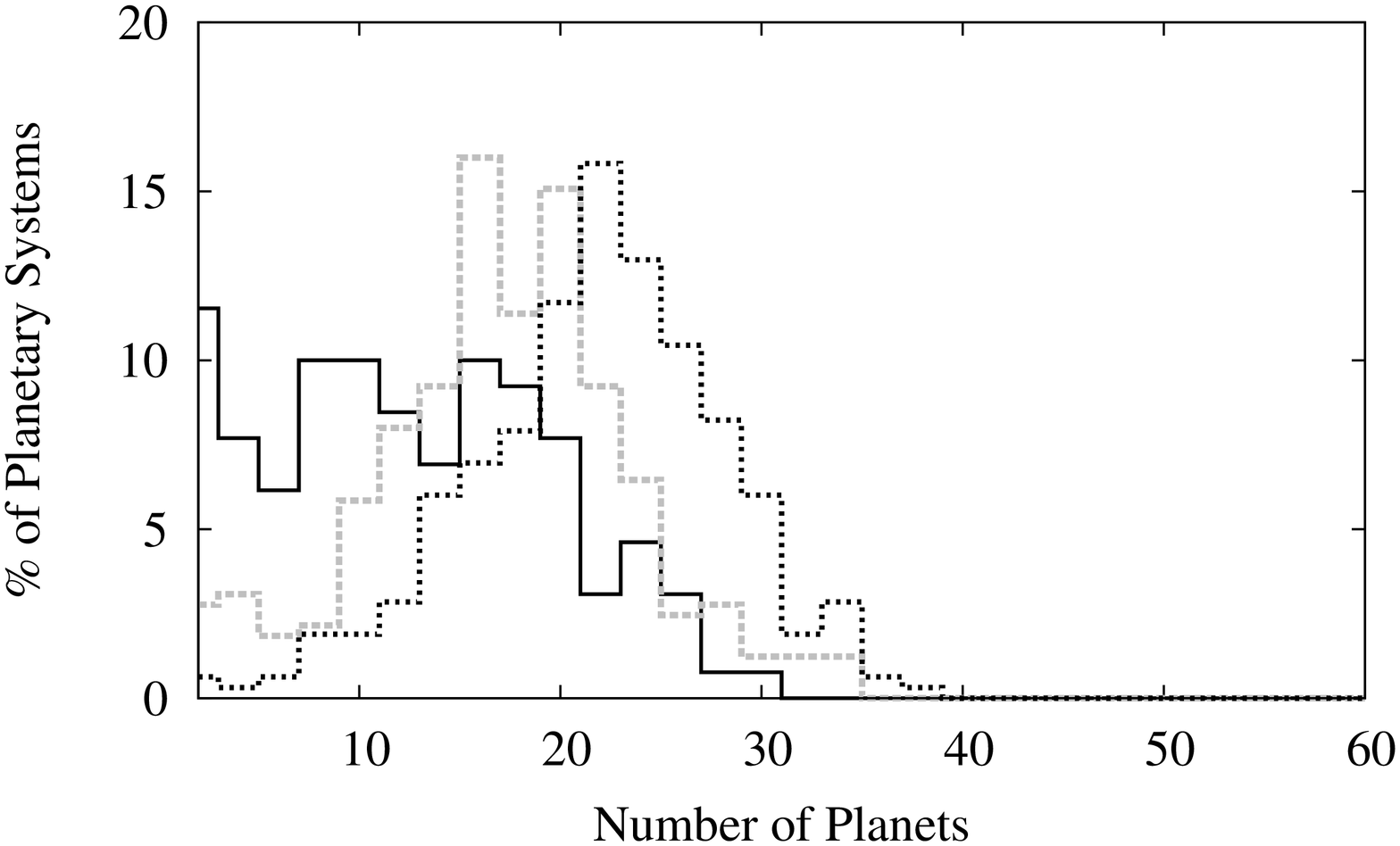}}
    \subfigure[]{\label{histo-pch-gama1}\includegraphics[angle=270,width=.5\textwidth]{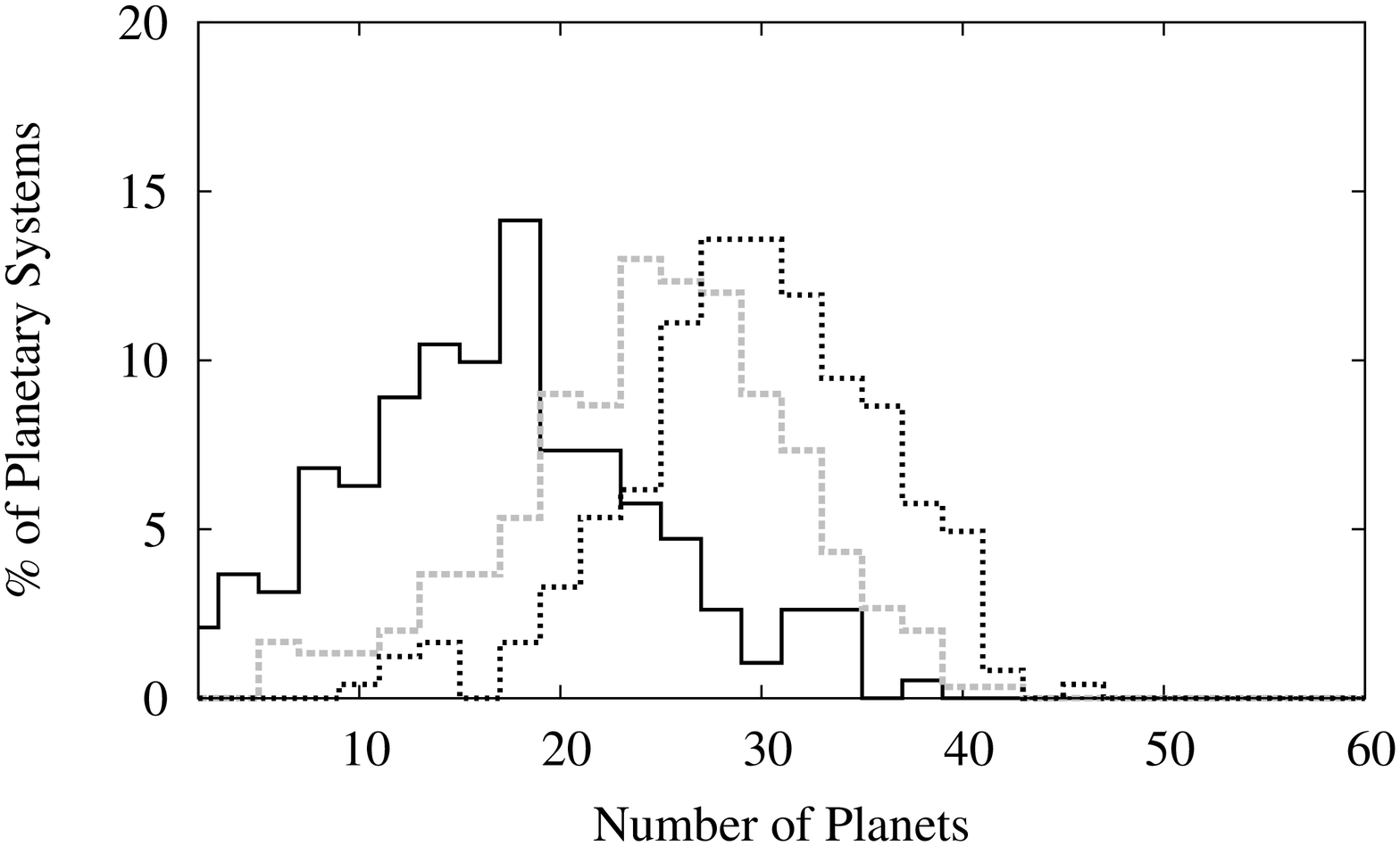}}
    \subfigure[]{\label{histo-pch-gama15}\includegraphics[angle=270,width=.5\textwidth]{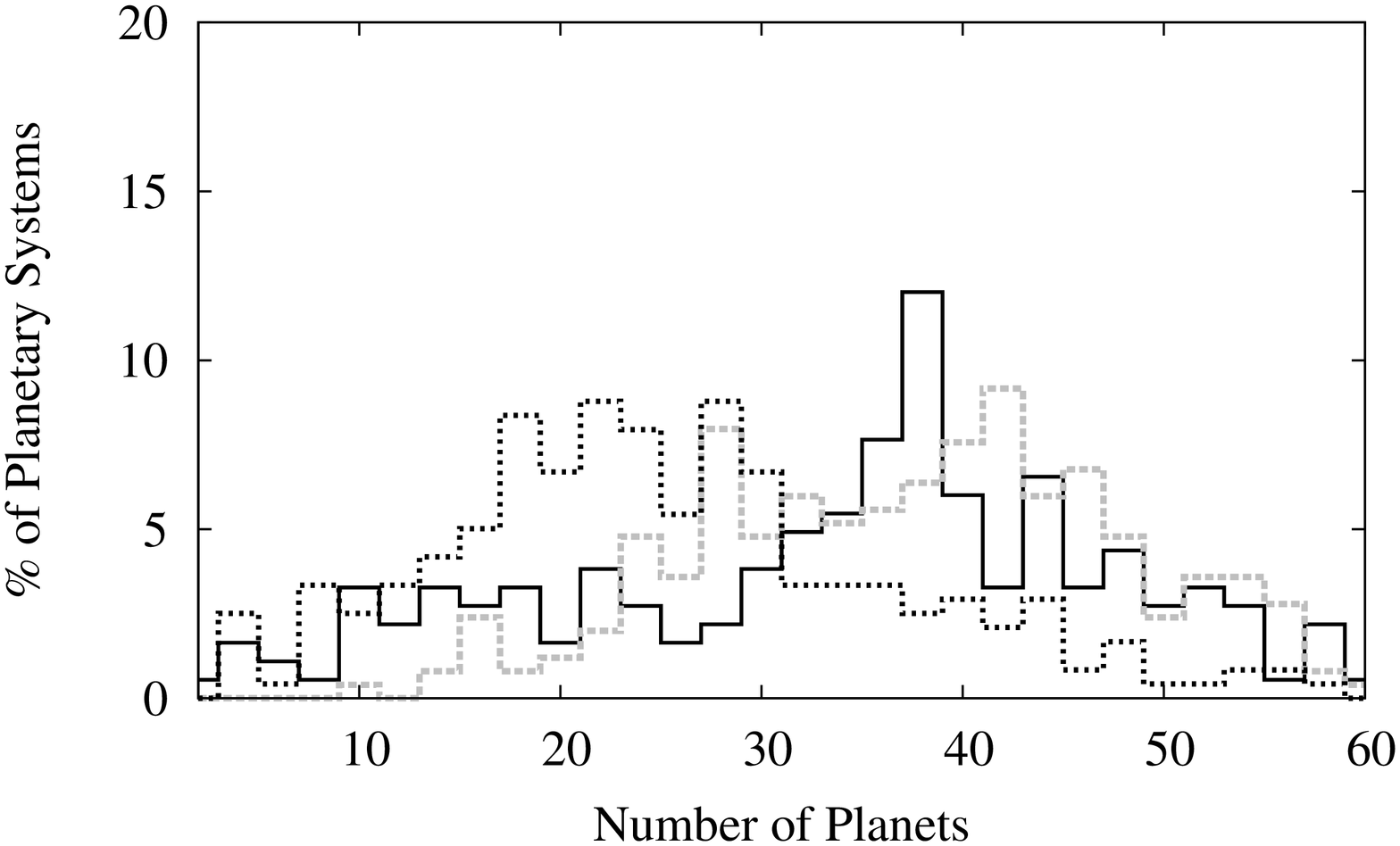}}
  \end{center}
  \caption{Histograms showing the number of planets per low mass planet system, where the planetary migration was not considered. Figure \ref{histo-pch-gama05} shows the results when we consider $\gamma=0.5$, figure \ref{histo-pch-gama1} shows the percentage of planetary systems with a certain number of planets when $\gamma= 1$ and finally the results obtained with $\gamma=1.5$ are shown in figure \ref{histo-pch-gama15}. In each figure, the solid line represents low mass planet systems formed in low mass discs, the gray dotted line shows the histogram for those formed in intermediate mass discs and the black dotted line shows the histogram for planetary systems formed from a massive disc.}
  \label{histo-pchicos-sinmig}
\end{figure}

Figures \ref{histo-pchicos-sinmig} show histograms representing the final number of embryos per planetary system. These plots where made based on the results found when the planetary migration is not considered. In the figures, the solid line shows the histogram for those planetary systems formed from a low mass disc, the gray dotted line represent the histogram for the systems that were born in an intermediate mass disc, and the black dotted line are the resulting histogram for very massive discs. As the planetary systems evolution is different when assuming different values for the $\gamma$ exponent, Figure \ref{histo-pch-gama05} shows the results when $\gamma=0.5$, Figure \ref{histo-pch-gama1} show the histograms resulting when $\gamma=1$ and when $\gamma=1.5$ we found the results shown in Figure \ref{histo-pch-gama15}.

The most massive discs have a lower
initial number of embryos because the separation between them is larger. This
causes that, in massive discs, embryos must be more massive in order to
collide with a neighbor core. When the disc profile is characterized by an
exponent in the inner disc given by $\gamma=0.5$, it is a rather flat disc and 
the growth slows much more on the inside than it increases on the outside, when compared to
discs with a larger $\gamma$. As a result the final number of embryos in these
discs is greater than the number reached in low mass discs, as could be seen in figure \ref{histo-pch-gama05}. We note in the Figure that most of the planetary systems formed in massive discs have between 20 and 25 final embryos, while those formed in low mass discs have between 10 and 20 rocky planets at the end of the simulation.  

When the profile is a bit steeper, $\gamma = 1$ this trend continues (Figure
\ref{histo-pch-gama1}), but when the disc is characterized by a density
profile of gas and solids with an exponent in the inner region of
$\gamma=1.5$, the growth of the embryos is really fast, so the merger between
the embryos is more frequent, and as a result the final number of embryos in
massive discs is smaller than the final number in low mass discs (Figure \ref{histo-pch-gama15}). 

\begin{figure*}
  \begin{minipage}{190mm}
    \begin{center}
      \includegraphics[angle=270,width=1.\textwidth]{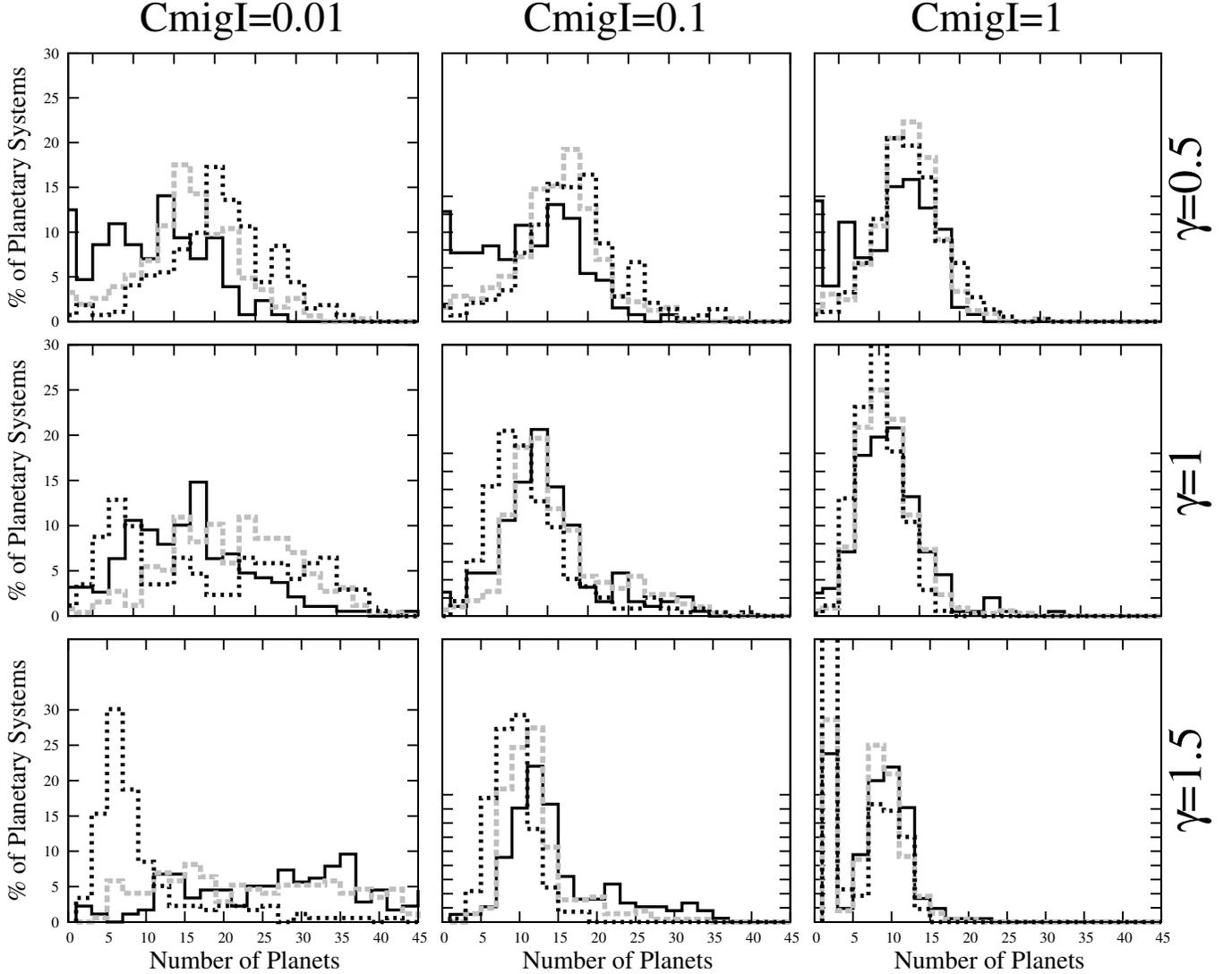}
    \end{center}
    \caption{The figure shows histograms which represent the number of planets harbored per low mass planet system. In each histogram, the simulation results are separated into three categories: planetary systems formed in low mass discs (solid line), planetary systems formed in intermediate-mass discs (gray dotted line) and planetary systems formed in massive discs (black dotted line). The rows show the results found when different initial disc profiles are assumed ($\gamma=0.5$, $1$ and $1.5$) and the columns show the numerical results when we assume different migration rates, $c_{migI}=0.01$, $0.1$ and $1$.}
    \label{histo-pchicos}
  \end{minipage}
\end{figure*}

As seen in figures \ref{histo-pch-gama05} , \ref{histo-pch-gama1} and \ref{histo-pch-gama15} the final number of embryos is strongly dependent on the initial disc profile. Nevertheless, this is not the only important factor, we are also interested in the effect of planetary migration in the final number of embryos. Figure \ref{histo-pchicos} shows which is the final number of planets per planetary system, when considering different migration rates and disc profiles. The columns represent different type I migration rates, where the first column shows the simulation results when $c_{migI}=0.01$, the second show the resulting histogram when the migration was delayed 10 times, and the last column represent the histograms when $c_{migI}=1$. The histograms in different rows, are the simulation results when different initial disc profiles are assumed (in the first $\gamma=0.5$, in the second $\gamma=1$ and in the last one $\gamma=1.5$). 

The interaction between an embryo and the surrounding gas, move the embryos towards the central star. This produces a higher number of collisions and the result of these mergers will be a smaller number of embryos at the end of the simulation. While this effect is important in determining the final number of embryos per planetary system, the migration rate depends on the disc profile, and as we note in previous figures, the separation between the embryos also depends on the value of $\gamma$ adopted, so the final number of embryos depends on the migration rate considered, but also on the initial disc profile. 

When considering a profile characterized by $\gamma=0.5$, we note that when $c_{migI}= 0.01$ the histogram is similar to the case without migration, although in this case the majority of planetary systems formed in very massive discs host $\sim20$ planets. When the migration is faster, $ c_ {migI} = 0.1 $, we note that the majority harbor between 15 and 20 final planets and when the migration is not delayed, most of planetary systems harbor less than 15 planets at the end of the simulation.

When the initial disc profile is characterized by $\gamma=1$ and $c_{migI}=
0.01$, we note that the majority of planetary systems formed from low massive
discs harbor between 15 and 20 final planets, while most of the low mass planet systems
formed from very massive discs end up with a number of planets between 5 and
10. When the migration rate is faster, the disc mass is less important and
most of the systems (formed from any disc), harbor less than 15 planets at the
end of the simulation. Finally, when the migration is not slowed down, we see
that the majority harbor less than 10 planets at the end of the simulation (20 million years).

The last row shows the histograms resulting when $\gamma=1.5$. In this case we
note that even a slow migration ($c_{migI}= 0.01$), cause that most of systems
formed from very massive discs harbor five planets at the end of the
simulation. On the other hand, the number of final planets formed in systems
with low mass discs, remains larger. When the migration is faster ($c_{migI}=
0.1$,), most of the planetary systems formed from any disc host 10 planets at
the end of the simulation. Finally, when $ c_{migI} = 1 $ we observe that most
of the systems formed in intermediate and low mass discs harbor between 5 and
10 planets while the vast majority of systems formed in very massive discs
harbor less than 5 planets at the end of the simulation. This is a consequence
of fast planetary migration. The migration is still more effective when the
disc mass and solid surface density is higher, which is the case of this last figure.

\subsection{Habitable Planets}

Considerations of stellar flux and planet climate lead to the definition of the habitable zone as the region where an Earth-like planet could support liquid water on its surface \citep{b29}. According to this definition, for the range of stellar masses considered in this work, the habitable zone lies between $0.9$ and $1.1~au$. But the evolution of Earth-like habitable planets is a complex process and locate a planet in the habitable zone is no guarantee for its habitability.

Although the potential habitability of an Earth-like planet depends on many factors as the tidal heating \citep{b32,b33}, its geophysical environment and atmospheric evolution \citep{b30}, the host star's activity and the planets' intrinsic magnetic field \citep{b31}, it is possible that planets in the habitable zone may contain water and host some life form. For this reason, missions that search for exoplanets intend to find small planets located in the habitable zone and the characterization of the stars and planetary systems that harbor them is important \citep{b38,b44}. 

On the other hand gas giant planets are far easier than terrestrial planets to
detect around other stars, and most of the planetary systems detected so far
are formed mostly by giant planets. Should we continue to monitor these
systems in the search for planets like Earth, or the presence of a gas giant
inhibits the formation of a small planet in the habitable zone?. When speaking
of hot and warm Jupiter systems, it is fairly clear that a migrating giant planet will cause any preexisting low-mass planets or planetesimals at smaller radii to be lost, either by accretion or scattering. What is not clear is whether a subsequent generation of planetesimals could form from the remnant disc after the giant path or if rocky planets formed farther out in the disc, subsequently migrate inward and locate at smaller radii and perhaps in the habitable zone.  

Recent studies dealing with these issues present varying results. On the one hand in \citet{b45}, he developed a simple model for the evolution of gas and solids in the disc and analyzed what remains after a massive planet migration. He found that the replenishment of solid material in the inner disc following giant planet inward migration, is generally inefficient and would not allow the formation of a second generation of habitable planets. On the other hand \citet{b46}, investigate the dynamics of post-migration planetary systems considering only one giant planet migrating and found that terrestrial accretion can occur during and after giant planet migration.  As seen this issue is not settled yet, and depends on many factors such as the time-scale of gas dissipation and the rates of accretion and planetary migration.

With our model we are not able to analyze the subsequent formation of
small planets, but we can analyze if it is possible that Earth-like
planets formed at a larger radii could end in the habitable zone due to a
subsequent migration. According to our model, when a giant planet migrates
towards the star the smaller embryos that found on its path are accreted or
scattered into external orbits. According to this, if we find an Earth-like
planet located in the habitable zone in a hot and warm Jupiter system, this means that
it was born at a larger radii (further than the giant planet formation zone) and a subsequent migration located it in the habitable zone.

Tables \ref{habitablesgama05}, \ref{habitablesgama1} and
\ref{habitablesgama15} show the percentage of planetary systems found in our
simulations that host planets woth masses less than $15~M_{\oplus}$, located between $0.9$ and $1.1~au$, in all the cases analyzed. 

\begin{table*}
 \centering
 \begin{minipage}{120mm}
\caption{Percentage (\%) of planetary systems formed when $\gamma=0.5$ and
  different migration rates, that host habitable planets.}
\begin{tabular}{lcccc}
\hline
\multicolumn{5}{|c|}{$\gamma=0.5$}\\
\hline
Type of Planetary System & No migration & $C_{migI}=0.01$  & $C_{migI}=0.1$  & $C_{migI}=1$ \\
\hline
\hline
Hot and warm Jupiters & 0 & 0 & 0 & 0.1 \\
Solar systems & 2.8 & 1.4 & 0.7 & 0.4 \\
Combined systems & 0 & 0 & 0 & 0 \\
Low mass planet systems & 2.5 & 0.5 & 1.9 & 8.1 \\
\hline
\label{habitablesgama05}
\end{tabular}
\end{minipage}
\end{table*}
\begin{table*}
 \centering
 \begin{minipage}{120mm}
\caption{Percentage (\%) of planetary systems formed when $\gamma=1$ and
  different migration rates, that host habitable planets.}
\begin{tabular}{lcccc}
\hline
\multicolumn{5}{|c|}{$\gamma=1$}\\
\hline
Type of Planetary System & No migration & $C_{migI}=0.01$  & $C_{migI}=0.1$  & $C_{migI}=1$ \\
\hline
\hline
Hot and warm Jupiters & 0 & 0 & 0.2 & 0.2 \\
Solar systems & 22.1 & 9.1 & 0.5 & 0 \\
Combined systems & 0 & 0 & 0 & 0 \\
Low mass planet systems & 38.7 & 19.7 & 14 & 9.5\\
\hline
\label{habitablesgama1}
\end{tabular}
\end{minipage}
\end{table*}
\begin{table*}
 \centering
 \begin{minipage}{120mm}
\caption{Percentage (\%) of planetary systems formed when $\gamma=1.5$ and
  different migration rates, that host habitable planets.}
\begin{tabular}{lcccc}
\hline
\multicolumn{5}{|c|}{$\gamma=1.5$}\\
\hline
Type of Planetary System & No migration & $C_{migI}=0.01$  & $C_{migI}=0.1$  & $C_{migI}=1$ \\
\hline
\hline
Hot and warm Jupiters & 0 & 0.4 & 0.3 & 0.2 \\
Solar systems & 22 & 3.4 & 0.4 & 0 \\
Combined systems& 0 & 0.1 & 0 & 0 \\
Low mass planet systems & 59.2 & 25.3 & 9.2 & 2.6\\
\hline
\label{habitablesgama15}
\end{tabular}
\end{minipage}
\end{table*}

As seen in tables, there are very few hot and warm Jupiter systems where a
subsequent migration locate low mass planets in the habitable zone. Since
the higher the density of gas in the disc, the faster the migration rate,
those discs with softer density profiles, require a faster migration rate in
order to move low mass planets in the habitable zone. For this reason, when
$\gamma=0.5$ habitable planets are only found in hot and warm Jupiter systems
when $c_{migI}=1$. When the disc density profile is sharper ($\gamma=1$), low mass planets in the habitable zone are found also when $c_{migI}=0.1$ and  when $\gamma=1.5$, there are planets who reach the habitable zone even for a migration rate as slow as the case which was delayed 100 times.

On the other hand we note that low mass, potentially habitable planets are found
preferably in planetary systems analogs to ours and also in low mass planet systems. These are the most favorable environments for the development of habitable planets.

\subsection{Mapping the Planetary System to its Birth Disc}\label{definiendo}

One of the key questions regarding planetary systems is how their properties reflect the conditions of their parent nebula. There are many key parameters which act for defining the architecture of a planetary system, here we explore the relevance of the initial disc mass, the characteristic radius, the metallicity, the time-scale of gas disc dissipation, the disc density distribution and migration rate. 

\subsubsection{Disc Mass and Characteristic Radius}

The mass of the disc and how it is distributed through the disc determines the
material available for the growth of embryos. A low mass disc will form low
mass planetary systems, while a massive disc will favor the formation of planetary systems with giant planets.  

The characteristic radius and exponent $\gamma$ of the density profile, are important because they say where most of the disc mass is distributed. A disc with a large $a_c$ and small value of $\gamma$, is more extended and allows the formation of cores further form the central star, but since the mass is distributed over a larger radius in a flat disc, this do not favor the mass concentration and therefore the frequency of giant planetary formation decreases.

\begin{figure*}
\begin{minipage}{190mm}
  \begin{center}
    \includegraphics[angle=270,width=1.\textwidth]{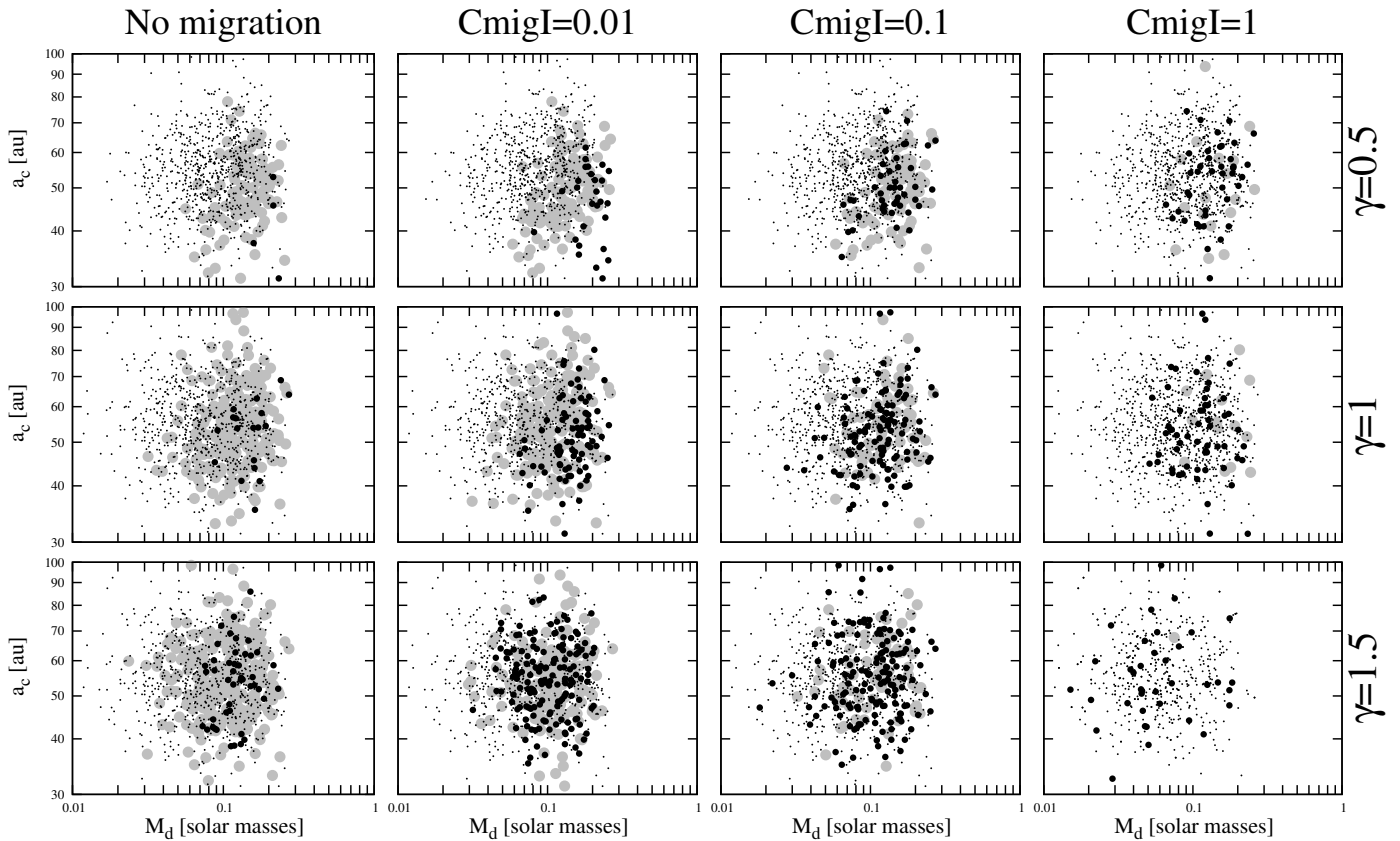}
  \end{center}
  \caption{In the figures each point represent a planetary system with a
    characteristic disc mass $M_d$ and $a_c$. Planetary systems analogs to our
    Solar System are shown in big gray dots, hot and warm Jupiter systems are
    the big black dots and the small black dots are low mass planet systems. Each row shows the results when different values for $\gamma$ are considered, in the first $\gamma=0.5$, in the second $\gamma=1$ and in the third row $\gamma=1.5$. The different columns represent different migration rates, being the first column the results when $c_{migI}=0$, in the second $c_{migI}=0.01$, in the third $c_{migI}=0.1$ and the last one shows the resulting planetary systems when the migration rate is not delayed ($c_{migI}=1$).}
  \label{md-rc}
\end{minipage}
\end{figure*}

Figure \ref{md-rc} shows the initial disc mass versus the characteristic
radius of the disc, where each point represents a planetary system and we show
the resulting planetary systems in all the cases considered in this work. The
gray big dots are those planetary systems analogs to our Solar System, the big
black dots are the hot and warm Jupiter systems and the small black dots show
the low mass planet systems. We noticed that both the characteristic radius and the
disc' s mass come from certain distributions, as was seen in section
\ref{resultados}. This fact is folded into the graphics and should be taken
into account.

According to our results shown in the figures, in the case of a disc with a
profile characterized by an exponent $\gamma=0.5$ on the inside (shown in the
first row of graphs), a mass of at least $0.06 M_{\odot}$ is needed to
allowing the formation of solar systems and their formation is favored when
considering a slow migration rate. We also note that in this case there are
very few hot and warm Jupiter systems when the migration is slow and it
increases when considering faster migration rates, but anyway,  mass of at
least $0.1 M_{\odot}$ is needed in order to allow the formation of this kind
of systems. We note, however, that in the case of a migration rate delayed
only 10 times, there are some systems that become hot and warm Jupiter systems with small initial discs, so a rapid migration rate allows the formation of these systems even for a relatively small $M_d$. It is also point out that in general, the formation of hot and warm Jupiters and solar systems occurs preferably with a quite small $a_c$, which is due to the fact that a lower $a_c$ favors the concentration of gas and solids and therefore the formation of giant planets.

In the second row we show the resulting planetary systems when $\gamma=1$ is
considered. In this case the disc present a larger density of solids and gas
in the inner disc and as result a mass of $\sim 0.04 M_{\odot}$ is enough to
form solar systems and there is no preferential $a_c$ to allow the formation
of these systems. A larger population of hot and warm Jupiter systems is found
even when the migration is not acting, which implies that there are some
initial discs which allow the formation in situ of these planets, but they are very rare.

Finally in the last row the planetary systems found when $\gamma=1.5$ are
shown, where we note that these discs, which present a high abundance of gas and
solids in the inner part, allow the formation of hot and warm Jupiters and solar systems even when the disc mass is $M_d \sim 0.02 M_{\odot}$ and for all $a_c$. 

In general, it can be noticed that the disc's mass has a huge influence on
defining the planetary system architecture, while the characteristics radius
of the disc is not generally a relevant factor. 
 
\subsubsection{Stellar Metallicity}

Precise spectroscopic studies biased toward high-metallicity stellar samples
\citep{b48}, those surveys who monitor stellar samples with low
metallicities \citet{b53} and other results with no bias \citet{b47}, have demonstrated that stars with giant planets tend to be particularly metal-rich when compared to the average local field dwarfs.

The physical mechanism for these correlation is of particular interest. One possible explanation could be that a high-metallicity protostellar cloud forms a metal-rich star and disc, and as an increased surface density of solids would facilitate the growth of embryonic cores then gas giant planet formation is greatly enhanced around more metal-rich stars. Alternatively, enhanced stellar metallicity may be a by-product of late-stage accretion of gas-depleted material and then a high metallicity star does not necessarily imply that the initial disc was rich in solids.

Most evidences today suggest that the metallicity excess has a "primordial'' origin \citep{b49,b50,b51,b52}, and for this reason we assume that metal-rich star implies metal-rich discs, and thus  the metal content of the cloud giving birth to the star and planetary system is indeed a key parameter to form a giant planet and determine the architecture of the planetary system.

\begin{figure*}
\begin{minipage}{190mm}
  \begin{center}
    \includegraphics[angle=270,width=1.\textwidth]{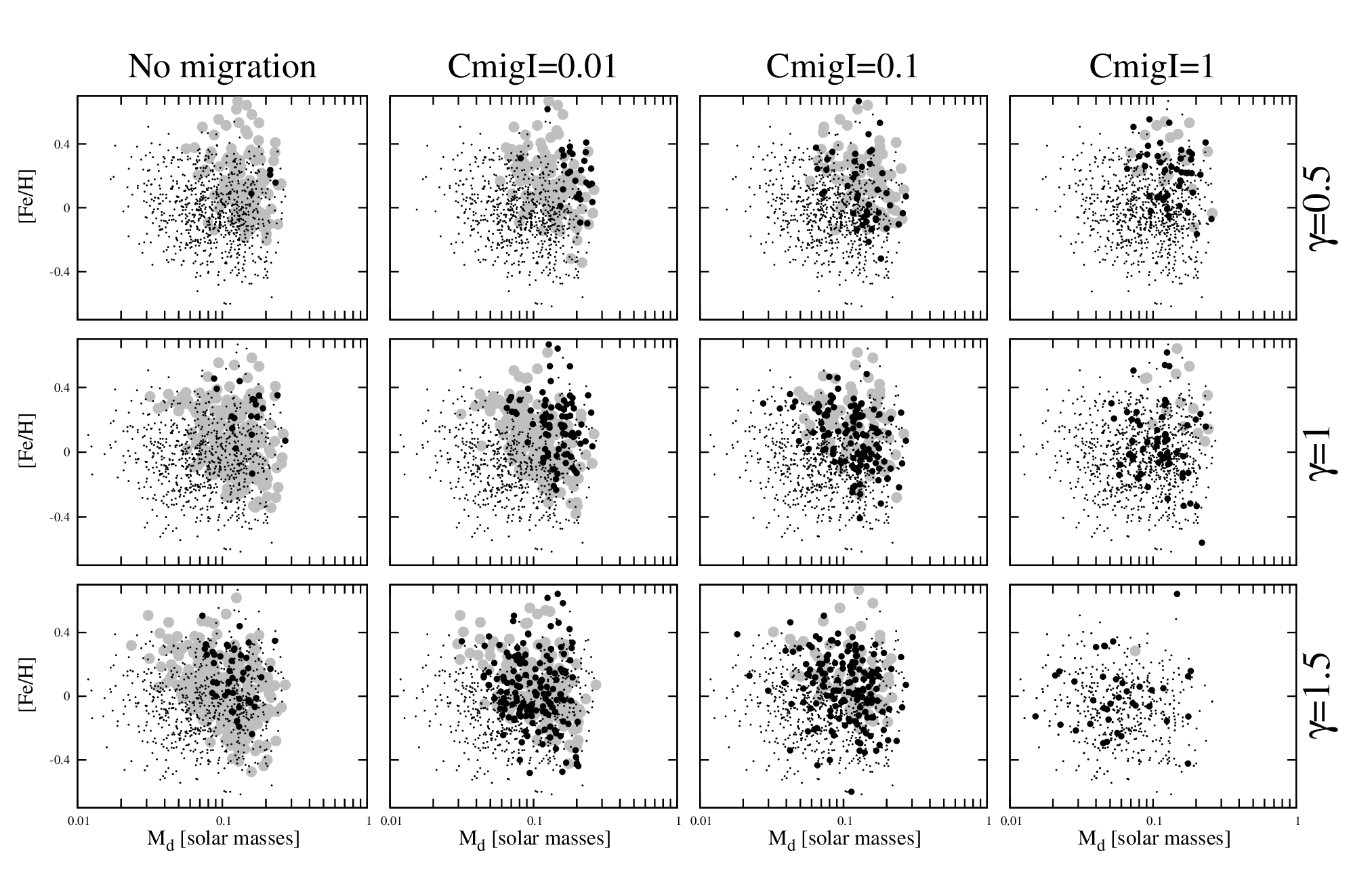}
  \end{center}
  \caption{The resulting planetary systems with a characteristic disc mass and
    metallicity are plot in the figures. Big gray dots represent the solar
    systems, the black big dots are hot and warm Jupiter systems and small
    black dots are low mass planet systems. Different columns shown the resulting planetary systems when different migration rates are considered: in the first column $c_{migI}=0$, in the second column the migration rate was delayed 100 times, in the third column $c_{migI}=0.1$ and in the last column $c_{migI}=1$. The different rows shows the results when the disc density has a profile characterized by different values of $\gamma$. Then the results found when $\gamma=0.5$ are shown in the firs row, in the second row $\gamma=1$ and in the last one $\gamma$ is equal to 1.5.}
  \label{md-metal}
\end{minipage}
\end{figure*}

\citet{b54} studied through numerical simulations the giant planet formation
in metal rich discs. They found that since the solid accretion rate increases
with the surface density of dust in the disc, the formation of gas giants tend
to be more prolific in a metal rich environment. Other authors find similar
results \citep{b99,b100}. Our results support this idea, but we also extend
this study to the role of high metallicities on the diversity of planetary
systems. In figure \ref{md-metal} each dot represent a planetary system (gray
dots are the solar systems, big black dots are hot and warm Jupiter systems
and small black dots are low mass planet systems), with a characteristic disc mass and stellar metallicity. The rows show the resulting planetary systems for the different density profiles considered in this work and the columns show the results with different migration rates.

As seen in the figures, when $\gamma=0.5$ (first row) a massive disc also needs
a metallicity of at east $-0.2$ in order to form solar systems and it has to
be larger than $-0.1$ to allow the formation of hot and warm Jupiter
systems. Then a massive disc is not the only important feature when forming
planetary systems with giant planets, a high metallicity disc is also
needed. There are planetary systems formed in massive discs, but as they are
characterized by low metallicities, they were not able to form giant planets
and remained as low mass planet systems.

When $\gamma=1$, the density profile is steeper and this allow the formation
of solar and hot and warm Jupiter systems at lower metallicities. In this case
the lower limit for solar systems formation is $-0.5$ and the formation of hot
and warm Jupiter systems is allowed when $[Fe/H] \ge -0.4$.

Finally in the last case, when $\gamma=1.5$, the solid surface density profile
of the disc are the sharpest, and this leads to a large concentration of
solids in the inner disc, which allows the formation of hot and warm Jupiter systems even for discs with metallicities of $-0.5$, while the lower limit for allowing the formation of solar systems is almost $-0.6$.    

A general result that we found, is that in most of the low metallicity discs,
low mass planet systems are the most common systems, independently on the disc
mass. This result could be in agreement with spectroscopic analysis found by
\citep{b51} and also with the results of surveys biased to low metallicity
stars, which do not found giant planets on stars with low metallicities
\citep{b53}. Nevertheless we note that these observational results could mean
that there are low mass planets around all these stars which were just not yet
detected (which is in agreement with our results) or that these stars have be no planets at all.

We also note that the higher the $\gamma$, the lower the
metallicity limit for allowing the formation of giant planets. As a
conclusion, a density profile characterised by $\gamma < 1.5$ is in better agreement with the observations.

\subsubsection{The Depletion of the Gas Disc}

Loss of the gaseous component of protoplanetary discs is due to a combination
of the accretion onto the central star \citep{b5}, photoevaporation
i.e. escape of the gaseous disc as a result of illumination by external
\citep{b105} or internal \citep{b106} ionizing radiation.  

According to observations in protoplanetary discs, the time-scales for the depletion of the gaseous component, ranges between 1 and 10 million of years \citep{b41,b42}. This time-scale is a key parameter in planetary systems formation because it sets a limit for the end of gas giant formation, affects the environment for terrestrial planet formation and determines the planetary migration as well. 

\begin{figure*}
\begin{minipage}{190mm}
  \begin{center}
    \includegraphics[angle=270,width=1.\textwidth]{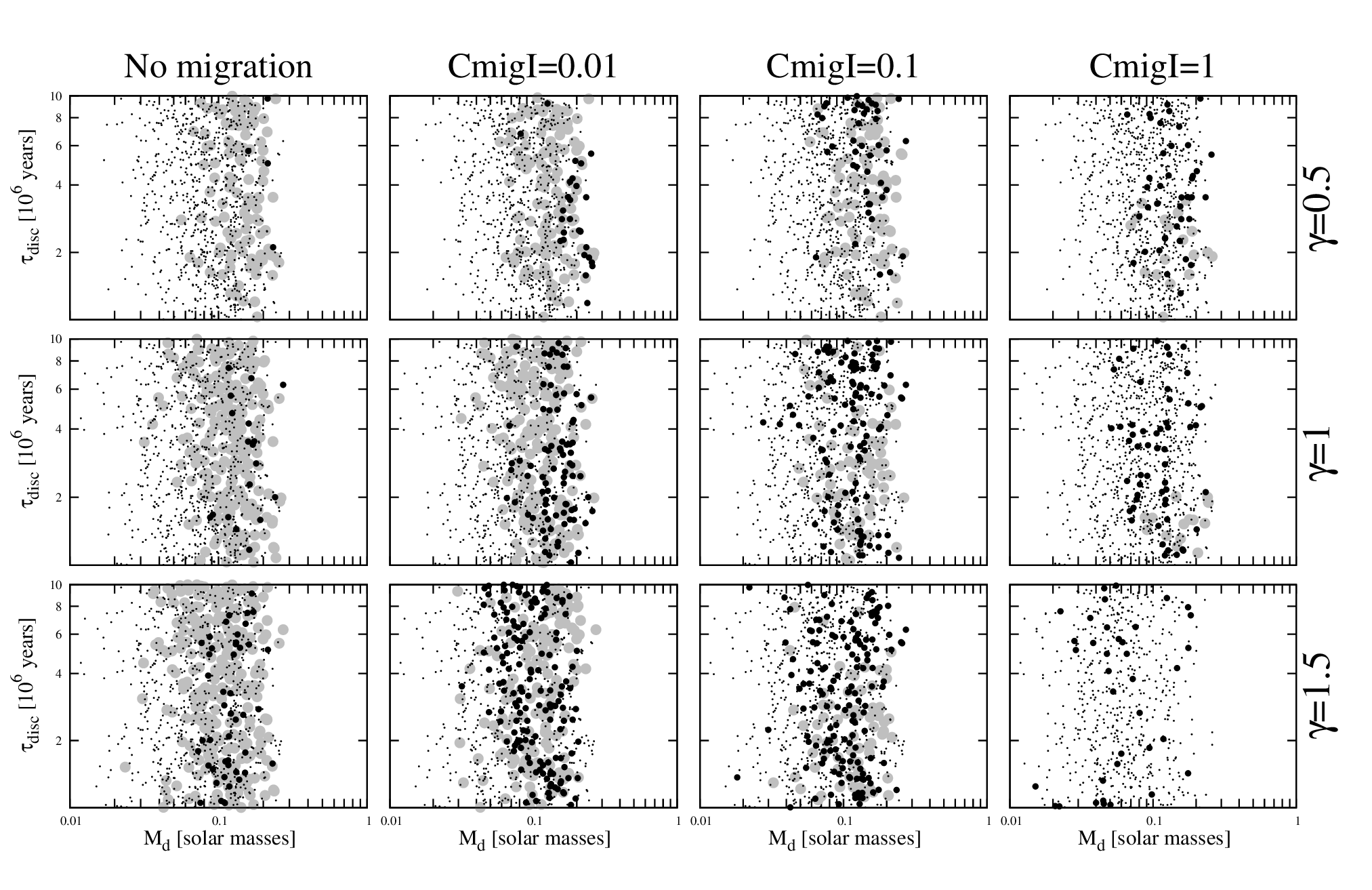}
  \end{center}
  \caption{The figure shows the dependence between the time-scale for the
    depletion of the gaseous disc and the $M_d$, for all the planetary systems
    formed. In this plot, the gray big dots represent the solar systems, the
    black big dots are the hot and warm Jupiter systems and the small black
    dots show low mass planet systems. The different columns represent the simulation results for different migration rates, being the first column the results when the migration was not considered, the second column shows the results when the migration rate is delayed 100 times, in the third column the type I migration was delayed 10 times and the last column show the results when the migracion is not delayed. The different rows represent the results when assuming different initial disc profiles: $\gamma=0.5$, $1$ and $1.5$, respectively.}
  \label{md-taugas}
\end{minipage}
\end{figure*}

Figure \ref{md-taugas} is an analogous to Figure \ref{md-metal}, but in this case the
gaseous disc characteristic time-scale vs disc mass is ploted.

We note that when the density profiles is $\gamma=0.5$ (first row), there is
no accumulation of solids in the disc, which leads to a lower accretion rate
and the time scale for the depletion of the gaseous disc is essential in
determining the architecture of the planetary system. As a result, we note
that a low disc mass combined with a faster depletion of the gaseous disc
leads to low mass planet systems. On the other hand a high disc mass combined with a
slow depletion of the gas leads to solar and hot and warm Jupiter systems.

When $\gamma=1$ and $\gamma=1.5$ (second and third row, respectively), the solid surface density profiles are steeper, as a result solids accumulate in the inner disc, promoting the rapid formation and migration of giant planets. Therefore, in this case of rapid formation, the gas dissipation timescale is not a relevant parameter in defining the architecture of the planetary system. 

These results are in agreement with those previously found by \citet{b56} which performed numerical simulations with a self consistent code with the aim of addressing how the properties of a mature planetary system map to those of its birth disk. To this end, they performed simulations covering a range of disc parameters and generated 100 planetary systems. Regarding the relevance of gaseous disc disipation time scale, they found similar results as we did, but with a more detailed code. 

\section{Summary and Conclusions}\label{conclusion}

The ensemble of more than 300 planetary systems discovered orbiting single stars, displays a wide range of architectures that show the planetary systems diversity. This diversity is related with the environment where the planets were formed and evolve. 

In order to study this diversity of extrasolar planetary systems, in the present work we have developed a semi-analytical code for computing planetary systems formation which is based on the core instability model for the gas accretion of the embryos and the oligarchic growth regime for the accretion of the solid cores. 

As our model is based on the core instability model, the mass distribution in the protoplanetary disc is important, since it defines the number and location of the final giant planets that would define the architecture of the planetary systems. Following protoplanetary discs observations, we explore different models for the initial protoplanetary nebula structure. Based on the similarity solutions for a viscous accretion disc, we assume that the gas and solid surface density are characterized by a power-law in the inner part of the disc, with an exponent $\gamma$ which take the values $\gamma=0.5$, $\gamma=1$ and $\gamma=1.5$, and an exponential decay in the outer part.

In our model we also assume that the embryos have an orbital evolution due to their interaction with the gaseous disc which leads to type I and II planetary migration. Type I is very fast and a factor for delaying this migration rate is assumed, in order to represent the effects that could slow down or even stop it. We assume that the migration rate is delayed 100 and 10 times and also analyze the cases where it is not delayed and in some simulations the migration is not considered. 

With this model we perform 12 simulations, in each one we explore the 3 different gas and solids disc density profiles considered and also the different planetary migration rates. In each simulation 1000 planetary systems are formed, whose initial conditions (mass and size of the disc, metallicity, mass of the central star and time-scale of gaseous disc dissipation) are taken randomly from distributions generated according to recent observational data. 

We analyze this artificial sample statistically, comparing with the observed
planetary systems. We also present a new classification of planetary systems
based on the location of giant planets and characterize each class exploring
how they reflect the disc where they were born, analyzing the importance of
key factors as disc size and mass, stellar metellicity, gas depletion
time-scale and planetary migration rate. We show the main characteristics of
each class and the number of giant planets that we expect to find in those
systems which harbor giant planets, and the final number of small planets
expected per low mass planet systems. Finally we analyze which are the best environment for the formation of small, potentially habitable planets and in which class of planetary system they are expected to be found.   

One of the striking results, is that in all cases analyzed it is always the
planetary systems with small planets which are the vast majority, being the
only planetary systems formed when considering low metallicity discs (when
comparing with the solar metallicity). Low mass planet systems are also the preferred planetary systems formed when a low mass disc is combined with a faster depletion of the gaseous disc, result also found by \citet{b56}. The final number of embryos per planetary system is strongly dependent on the initial disc profile and migration rate assumed. 

Planetary systems analogous to our Solar System are preferentially formed in
massive discs, in a high metallicity environment and where the disc profile
that defines the gas and solid density in the inner disc is small. In
addition, it also requires that the migration rate is not too fast, in order
to avoid that the planets are pushed toward the inner edge of the disc. We
found solar systems with more than one giant planet, even with three, but in
no case with more than three giant planets. 

Assuming the sharpest disc profile, characterized by the largest value of
$\gamma$ assumed in this work, implies that the solids are accumulated in the
inner regions of the disc, while they fall rapidly beyond the characteristic
radius of the disc. This accumulation in the inner disc, favors the formation
of hot-Jupiter planets, which are preferably formed in very massive discs
combined with a slow depletion of the gas and a metal rich environment. Also a
fast migration rate is required in order to form these systems. According to
our results, most of the hot and warm Jupiter systems are composed by only one giant planet, which is also a tendency of the current observational data. 

We do not found any planetary system with giant planets located further from
$30~au$, which is a consequence of the scenario assumed for giant planet formation.

We also analyze which are the most favorable environments for the formation of
low mass, potentially habitable planets and found that they are preferably formed
in planetary systems analogs to our Solar System and also in low mass planet systems, which are the best environments for the developing of these systems.

\end{document}